%% file: paperdraft.tex
\documentclass[a4paper,11pt]{article}

\usepackage{latexsym}
\usepackage{a4wide}
\usepackage{graphicx, subfigure}
\usepackage{cite}
\usepackage{tabularx}
\usepackage{array}
\usepackage{graphics}     
\usepackage{amsmath}
\usepackage{amssymb} 
\usepackage{longtable} 
\usepackage{verbatim} 
\usepackage[table]{xcolor}
\usepackage{hyperref} 
\usepackage{mathtools}
\usepackage{soul}

\definecolor{light-gray}{gray}{0.90}
\definecolor{light-yellow}{RGB}{255, 255, 128}
    
\definecolor{mygreen}{RGB}{143, 176, 50}
\definecolor{myorange}{RGB}{225, 156, 36}
\definecolor{myblue}{RGB}{94, 129, 181}

\renewcommand{\arraystretch}{1.2}

\newcommand{\ra}[1]{\renewcommand{\arraystretch}{#1}}
\newcommand{\rb}[1]{\renewcommand{\tabcolsep}{#1}}

\begin{document}

\hfill {\tt CERN-TH-2017-031, IPM/PA-455, MITP/17-007}  

\def\thefootnote{\fnsymbol{footnote}}
 
\begin{center}

\vspace{2.5cm}  

{\huge\bf {Large hadronic power corrections or\\ new physics in  the rare decay $B \to K^* \ell\ell$ ?} }

\setlength{\textwidth}{11cm}
                    
\vspace{2.cm}
{\Large\bf
V.G.~Chobanova\footnote{Email: veronika.chobanova@cern.ch}$^{,a}$,
T.~Hurth\footnote{Email: tobias.hurth@cern.ch}$^{,b}$,
F.~Mahmoudi\footnote{Also Institut Universitaire de France, 103 boulevard Saint-Michel, 75005 Paris, France\\ \hspace*{0.49cm} Email: nazila@cern.ch}$^{,c,d}$,\\
D.~Mart\'inez Santos\footnote{Email: Diego.Martinez.Santos@cern.ch}$^{,a}$,
S.~Neshatpour\footnote{Email: neshatpour@ipm.ir }$^{,e}$
}
 
\vspace{1.cm}
{\em $^a$Instituto Galego de F\'isica de Altas Enerx\'ias, Universidade de Santiago de Compotela, Spain}\\[0.2cm]
{\em $^b$PRISMA Cluster of Excellence and  Institute for Physics (THEP)\\
Johannes Gutenberg University, D-55099 Mainz, Germany}\\[0.2cm]
{\em $^c$Univ Lyon, Univ Lyon 1, ENS de Lyon, CNRS, Centre de Recherche Astrophysique de Lyon UMR5574, F-69230 Saint-Genis-Laval, France}\\[0.2cm]
{\em $^d$Theoretical Physics Department, CERN, CH-1211 Geneva 23, Switzerland}\\[0.2cm]
{\em $^e$School of Particles and Accelerators, Institute for Research in Fundamental Sciences (IPM)
P.O. Box 19395-5531, Tehran, Iran}\\[0.2cm]

\end{center}

\renewcommand{\thefootnote}{\arabic{footnote}}
\setcounter{footnote}{0}

\vspace{1.cm}
\thispagestyle{empty}
\centerline{\bf ABSTRACT}
\vspace{0.5cm}
The well-known problem of the unknown power corrections within QCD improved factorisation leaves the interpretation  of the so-called LHCb anomalies in the angular observables of $B \to K^{*} \mu^+\mu^-$ as an open problem. In order to contribute to the question  whether they represent a first sign for new physics beyond the Standard Model or a consequence of underestimated hadronic power corrections, we present a direct comparison of two global fits to the data based on the two different assumptions. In addition, we summarise the possible options  to resolve this puzzle in the future. 

\newpage


\input{introduction}
\input{theory}
\input{results}
\input{results_details}

\input{results_Wilktests1}

\input{results_Wilktests2}
\input{prospects}

\input{conclusions}


\section*{Acknowledgement} 
TH thanks the CERN theory group for its hospitality during his regular visits to CERN where part of this work was written.
SN thanks the Mainz Institute for Theoretical Physics (MITP) for hospitality while part of this work was being completed.
The authors are grateful to  J.~Camalich for useful discussions.


\newpage
\appendix
\input{constrained}


\newpage

\providecommand{\href}[2]{#2}\begingroup\raggedright\endgroup

\end{document}

%% file: introduction.tex
\section{Introduction}\label{sec:intro}

In 2013, the LHCb collaboration presented the full angular analysis of the $B \to K^* \mu^+ \mu^-$ observables with 1 fb$^{-1}$ of data~\cite{Aaij:2013qta}.
The experimental measurement of most of these observables was in good agreement with the Standard Model predictions.
However, there were some deviations from the Standard Model (SM) predictions in certain bins of the dilepton invariant mass ($q^2$) for some of the observables where the largest deviation with 3.7$\sigma$ significance was reported in the $q^2 \in [4.30,8.68]$ GeV$^2$ bin for the  angular observable $P_5^\prime$.
These deviations in the angular observables were reaffirmed by the LHCb collaboration with the 3 fb$^{-1}$ dataset~\cite{Aaij:2015oid}, in the smaller $[4.0,6.0]$ and $[6.0,8.0]$ GeV$^2$ bins. Moreover, another deviation in the branching ratio of the decay $B_s\to \phi \mu^+ \mu^-$ was found by LHCb~\cite{Aaij:2015esa}, where the experimental results are below the SM predictions. 
And recently, the Belle collaboration also reported\cite{Abdesselam:2016llu} a deviation for $P_5^\prime$ 
in the $[4,8]$ GeV$^2$ bin with $2.1\sigma$ significance where the experimental uncertainty is larger than that of LHCb.

Assuming these anomalies to be due to some New Physics (NP) contribution, global analyses of the $b\to s$ data showed that all these deviations can be best explained by NP contributions to the Wilson coefficient 
$C_9$~\cite{Descotes-Genon:2013wba,Altmannshofer:2013foa,Beaujean:2013soa,Horgan:2013pva,Hurth:2013ssa,Mahmoudi:2014mja}.
On the other hand, the anomalies in the $B\to K^* \mu^+ \mu^-$ data  can also be explained by underestimated hadronic effects arising from power corrections~\cite{Jager:2012uw,Jager:2014rwa,Ciuchini:2015qxb}.

The standard theory framework for these exclusive decay modes in the low-$q^2$ region is QCD factorisation (QCDf) and its field theoretical formulation Soft-Collinear Effective Theory (SCET).
It is well-known that there is  no theoretical description of power corrections existing 
within this theoretical framework. Power corrections can only be {\it guesstimated}  and are -- depending on the involved assumptions -- rather different in size~\cite{Jager:2012uw,Jager:2014rwa,Descotes-Genon:2014uoa,Ciuchini:2015qxb}. 
Thus, it is difficult to unambiguously determine whether the source of the anomaly is due to NP or due to underestimated hadronic effects within these exclusive observables. 
In fact, we have illustrated in Ref.~\cite{Hurth:2016fbr,Mahmoudi:2016mgr} how the significance of the anomalies depend on the guesstimate of the non-factorisable  power corrections. Moreover, we showed~\cite{Hurth:2016fbr,Mahmoudi:2016mgr} that the non-factorisable power corrections required to explain the disagreement between SM predictions and experimental measurements are very large compared to the leading non-factorisable piece of the QCDf amplitude. In some critical bins this amounts to larger than 150\% corrections compared to the leading non-factorisable contributions of QCDf at the \emph{amplitude} level which somewhat questions the validity of the QCDf approach.

In case the puzzle of the LHCb anomalies remains unresolved for some time, the Belle II results for the $B\to X_s \ell^+ \ell^-$ decay can  determine the source of the anomaly  as explicitly shown 
in Refs.~\cite{Hurth:2013ssa,Hurth:2014zja,Hurth:2014vma,Hurth:2016fbr}. In contrary to the exclusive decay, for the inclusive case, the power corrections can be theoretically estimated (see Refs.~\cite{Hurth:2010tk,Hurth:2007xa,Hurth:2003vb} for reviews). However, there are options to resolve the puzzle  before Belle II:

LHCb has presented another $2.6\sigma$ deviation in the ratio  $R_K\equiv {\rm BR}(B\to K^+ \mu^+ \mu^-)/{\rm BR}(B\to K^+ e^+ e^-)$~\cite{Aaij:2014ora}. The SM prediction of $R_K$ is quite precise and free of large theoretical uncertainties.
So this deviation cannot be explained by power corrections. If the experimental result is reconfirmed with future measurements, flavour violating NP contributions would be the most probable explanation.
Interestingly, this  $2.6\sigma$ deviation can also be explained with a similar NP contribution to 
$C_9$ like the anomalies in the angular observables~\cite{Alonso:2014csa,Hiller:2014yaa,Ghosh:2014awa,Hurth:2014vma,Altmannshofer:2014rta,Straub:2015ica,Altmannshofer:2015sma,Hurth:2016fbr}.
Thus, a confirmation of the deviation in $R_K$ would also indirectly confirm the NP interpretation of the anomalies in the angular observables. 
Updated measurement of $R_K$ with a larger dataset as well as other theoretically clean ratios of 
$b \to s \ell^+ \ell^-$ observables which test lepton universality~\cite{Altmannshofer:2014rta,Hurth:2016fbr, Capdevila:2016ivx,Serra:2016ivr} would be illuminating in this regard.

A theoretical estimation of the power corrections is possible through Light-Cone Sum Rule (LCSR) approach for small  $q^2$ which can be extrapolated to higher $q^2$ close to the charmonium resonances
via dispersion relations and  a phenomenological model~\cite{Khodjamirian:2010vf,Khodjamirian:2012rm}. 
But  for the $B \to K^* \mu^+ \mu^-$ decay only a partial estimate for small $q^2$  is available~\cite{Khodjamirian:2012rm}. 
Future estimation of the power corrections through the LCSR approach  might allow to establish or to disprove NP in the angular observables in $B\to K^* \mu^+ \mu^-$.

In this paper we summarise the results which were presented at the workshop ``Implications of LHCb measurements and future prospects'', CERN, Geneva, 12-14 October 2016 ~\cite{Nazilatalk:2016}. 
We follow another (modest) approach and directly compare the fit of possible unknown 
non-factorisable power corrections with a fit to NP  by statistical methods.
We already anticipate that such a comparison of different fits to the present data only offers hints to possible resolutions of the puzzle of the LHCb anomalies but does not allow to resolve it at present. 

The paper is organised as follows. In section~\ref{sec:theory} we give an overview of the various hadronic contributions that are relevant for the $B\to K^* \mu^+ \mu^-$ decay and discuss how potential non-factorisable power corrections could appear in terms of the helicity amplitudes and compare it to the NP contributions which can have a similar effect. In section~\ref{subsec:details} details of our fits are discussed. In section~\ref{subsec:Wilktest1} we give the result of the fit to NP in the Wilson coefficients $C_7$ and/or $C_9$ as well as of the fit to unknown power corrections. 
We statistically compare the fits via likelihood ratio tests. In section~\ref{subsec:Wilktest2} we also consider  possible NP in the Wilson coefficient $C_{10}$ in our analysis. 
In section~\ref{subsec:prospects} we discuss the prospects of the LHCb upgrade.
We conclude in section~\ref{sec:conclusions}.

%% file: theory.tex
\section{Theoretical Setup}\label{sec:theory}
The effective Hamiltonian describing $b \to s \ell^+ \ell^-$ processes can be written as a sum
of a hadronic and a semileptonic part (see Ref.~\cite{Jager:2012uw}),
\begin{align}
 {\cal H}_{\rm eff} = {\cal H}_{\rm eff}^{\rm had} + {\cal H}_{\rm eff}^{\rm sl}\,,
\end{align}
with
\begin{align}
 {\cal H}_{\rm eff}^{\rm had}=-\frac{4G_F}{\sqrt{2}}V_{tb}V_{ts}^*\sum_{i=1,\ldots,6,8}C_i\; O_i\,, 
 && {\cal H}_{\rm eff}^{\rm sl}=-\frac{4G_F}{\sqrt{2}}V_{tb}V_{ts}^*
 \sum_{\mathclap{\substack{i=7,9,10,\\
                   S,P,T}}}(C_i\; O_i + C^\prime_i\; O^\prime_i)\, ,
\end{align}
where $C_{i}$ denote the Wilson coefficients. The explicit form of the effective operators $O_{i}$ which we use in this paper is given in Ref.~\cite{Chetyrkin:1996vx}.

For the exclusive $B\to K^* \mu^+ \mu^-$ decay the dominant contribution of the effective Hamiltonian
is from the semileptonic part, ${\cal H}_{\rm eff}^{\rm sl}$. 
The hadronic matrix element of ${\cal H}_{\rm eff}^{\rm sl}$ can be parametrised in terms of 
seven independent form factors which are calculated via methods such as LCSR or lattice QCD.
Due to the non-perturbative nature of the forces pertinent to these calculations,
the form factors are usually considered as one of the main sources of theoretical uncertainty.
At large energy and in the heavy quark limit, employing the kinematic constraints and 
the emerging symmetry relations, the $B \to K^*$ form factors can be
written in terms of only two soft form factors, $\xi_\perp$ and $\xi_\parallel$,
up to correction of $\alpha_s$ and $\Lambda/m_{b}$. 
Hence, at leading order, form factor independent observables can be constructed, reducing the 
theoretical uncertainty.
However, there are symmetry breaking corrections of ${\cal O}(\alpha_s)$ and ${\cal O}(\Lambda/m_{b})$ to the relations among the seven full form factors and the two soft form factors,
and while the former corrections have been calculated within the QCDf framework~\cite{Beneke:2000wa} 
the beyond leading $\Lambda/m_b$ powers, referred to as {\it factorisable power corrections}, are unknown in the first place.
They introduce a new source of theoretical uncertainty~\footnote{ As was advocated first in Ref.\cite{Jager:2012uw} (see also Ref.~\cite{Descotes-Genon:2014uoa}), the factorisable power corrections can be in  principle determined using the factorisation formula between soft and full form factors which can be written  in a schematic way:
\begin{equation} \label{FACT2}
F_{\rm full} (q^2)  = D \xi_{\rm soft} + \Phi_B \otimes T_F \otimes \Phi_M  + {\cal O}(\Lambda_{\rm QCD}/m_b)\,,
\end{equation}
where $D$ and $T_F$ are perturbatively calculable functions. There are non-factorisable contributions (second term), but also power corrections (third term) which are the factorisable power corrections to be determined by this equation. It is clear that within this determination the uncertainties of the LCSR calculations of the QCD form factors are transmitted to the factorisable power corrections.}. 
It is possible to avoid these factorisable power corrections entirely by employing the full form factors instead 
of the two soft form factors.
Recently new  LCSR results on the $B\to K^*$ form factors, which also include the correlations among
the form factor uncertainties have been presented~\cite{Straub:2015ica}. 

Even if the $B\to K^*$ form factors were precisely known, there would still be another source of hadronic uncertainty.
Besides the semi-leptonic part of the effective Hamiltonian, the hadronic part also contributes to the $B\to K^* \mu^+ \mu^-$ decay through the emission of a virtual photon which decays into a lepton pair.
These non-local effects are expressed through the $B \to K^*$ matrix elements of the time ordered products of 
the electromagnetic currents~$j_{{\rm em},{\mu}}^{\rm had/lept}$, and the hadronic effective Hamiltonian
\begin{align}\label{eq:hadAmp}
\mathcal A^{\rm (had)} = &- i \frac{e^2}{q^2} \!\! \int \!\! d^4x e^{- i q \cdot x}
   \langle \ell^+ \ell^- | j_{{\rm em},\mu}^{\rm lept}(x) | 0 \rangle 
 \times \!\! \int \!\! d^4 y\, e^{i q \cdot y}   \langle \bar K^* | T \{ j_{\rm em}^{\rm had, \mu}(y)
 {\mathcal H}^{\rm had}_{\rm eff}(0) \} | \bar B \rangle\,.
\end{align}
The one-loop contributions of the four-quark operators can be described through the matrix element of $O_9$ and hence are usually taken into account in the form of corrections to $C_9$ via the effective Wilson coefficient 
$C_9^{\rm eff}(=C_9+Y(q^2))$~\cite{Grinstein:1988me,Misiak:1992bc,Buras:1994dj}.
There are other contributions from ${\cal H}_{\rm eff}^{\rm had}$ such as weak annihilation and soft gluon exchange~\cite{Beneke:2001at} which are more complicated to estimate 
and cannot be factorised into form factors and leptonic currents.
These contributions which are referred to as non-factorisable corrections
can be treated at large recoil energy within the QCD factorisation framework where
they are factorised as a convolution of $B$ and $K^*$ 
distribution amplitudes with hard-scattering kernels.
The QCDf calculations are applicable below the charmonium resonances and are available at leading power in 
$\Lambda/m_b$~\cite{Beneke:2001at,Beneke:2004dp}.
Nevertheless, higher powers of the non-factorisable contributions (referred to as {\it non-factorisable power corrections}) remain unknown.
There is no theoretical description of such power corrections existing within the QCDf approach, but there
are some partial calculations of these power corrections within the LCSR approach available~\cite{Khodjamirian:2010vf} as already discussed in the introduction.

The contributions from  the hadronic part of the effective Hamiltonian 
can be conveniently described via helicity amplitudes. In the SM we have three non-trivial amplitudes (for the
general case we refer the reader to Ref.~\cite{Jager:2012uw}):  
\begin{align}\label{eq:HV}     
  H_V(\lambda) &=-i\, N^\prime \Big\{ C_{9}^{\rm eff} \tilde{V}_{\lambda}(q^2) 
      + \frac{m_B^2}{q^2} \Big[\frac{2\,\hat m_b}{m_B} C_{7}^{\rm eff} \tilde{T}_{\lambda}(q^2)
      - 16 \pi^2 {\cal N}_\lambda(q^2) \Big] \Big\}\, , \\
  H_A(\lambda) &= -i\, N^\prime C_{10}  \tilde{V}_{\lambda} \, , \\
  H_P &= i\, N^\prime \Big\{  \frac{2\,m_\ell \hat m_b}{q^2} 
                     C_{10}( 1 + \frac{m_s}{m_b} )\tilde{S}\Big\}\,,
\end{align}
where $\lambda=\pm 1,0$ represent the helicities.
The contributions of  ${\cal H}_{\rm eff}^{\rm had}$ to the $B\to K^* \ell^+ \ell^-$ decay
takes place through the emission of a virtual photon which involves an electromagnetic current which is vectorial.
Thus, all these effects are accounted for in the vector helicity amplitudes $H_V(\lambda)$, via the effective part of $C_9$ and ${\cal N}_\lambda$ which stands for the non-factorisable contributions.
The latter can be decomposed in a leading part which can be calculated in QCDf and into the non-factorisable power
corrections denoted here as $h_\lambda(q^2)$ 
\begin{align}\label{eq:Nlambda}
 {\cal N}_\lambda (q^2) \equiv \text{Leading order in QCDf} + h_\lambda (q^2)\,.
\end{align}

The power corrections $h_\lambda$, for which no complete estimation is available can be fitted to data as was shown explicitly in Ref.~\cite{Ciuchini:2015qxb}. 
Following the parametrisation of that analysis we take $h_{\lambda}(q^2)$ to be
\begin{align}\label{eq:hlambda}
  h_\lambda(q^2)= h_\lambda^{(0)} + \frac{q^2}{1 {\rm GeV}^2}h_\lambda^{(1)} 
  + \frac{q^4}{1 {\rm GeV}^4}h_\lambda^{(2)}\,,
\end{align}
where $h_\lambda^{(0,1,2)}$ can each be a complex number resulting in overall 18 free real parameters.
Considering Eq.~(\ref{eq:HV}), the effect of the hadronic non-factorisable power corrections on the helicity amplitude is written as
\begin{align}\label{eq:deltaHVPC}
\delta H_V^{\lambda,{\rm PC}}&=  i N^\prime m_B^2 \frac{16 \pi^2}{q^2}
 h_\lambda(q^2) = i N^\prime m_B^2 \frac{16 \pi^2}{q^2}
\left( h_\lambda^{(0)}+q^2 h_\lambda^{(1)}+q^4 h_\lambda^{(2)} \right)\,.
\end{align}
We note that there is a $1/q^2$ term multiplying $h_\lambda(q^2)$ 
resulting in an overall $q^2$ expansion with terms  $1/q^2, 1$ and $q^2$.
From Eq.(\ref{eq:HV}), it is clear that any contribution to $C_7$ and/or $C_9$ can be interpreted as power corrections.
Incidentally, new physics scenarios which explain the tensions between SM predictions and experimental results,
show preference for a reduction in $C_9$ which can be mimicked by hadronic corrections and
makes it  difficult to identify the source of the tension.

\begin{figure}[t!]
\centering
\includegraphics[width=0.49\textwidth]{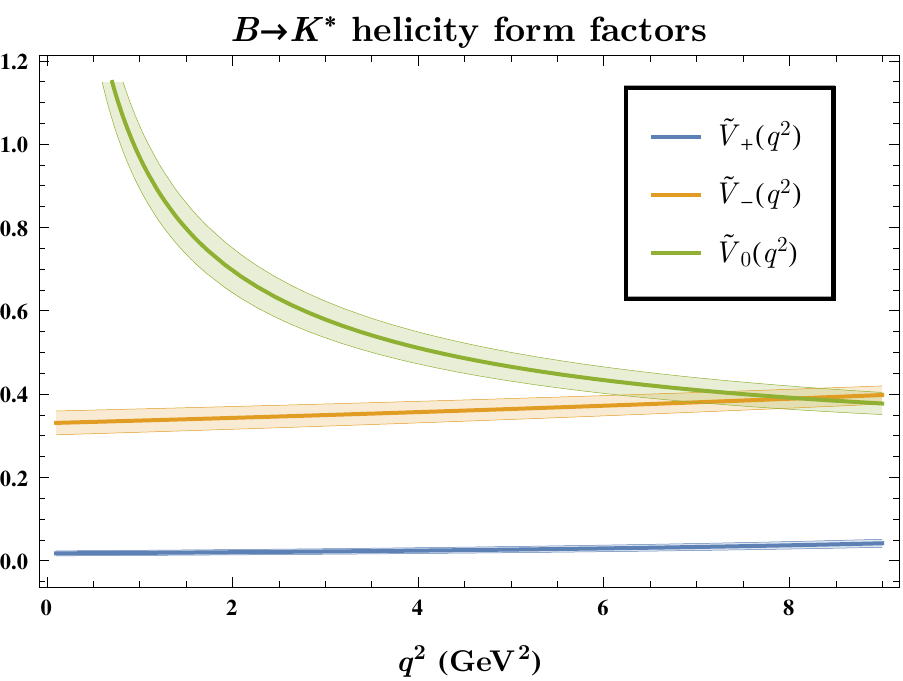}
\includegraphics[width=0.49\textwidth]{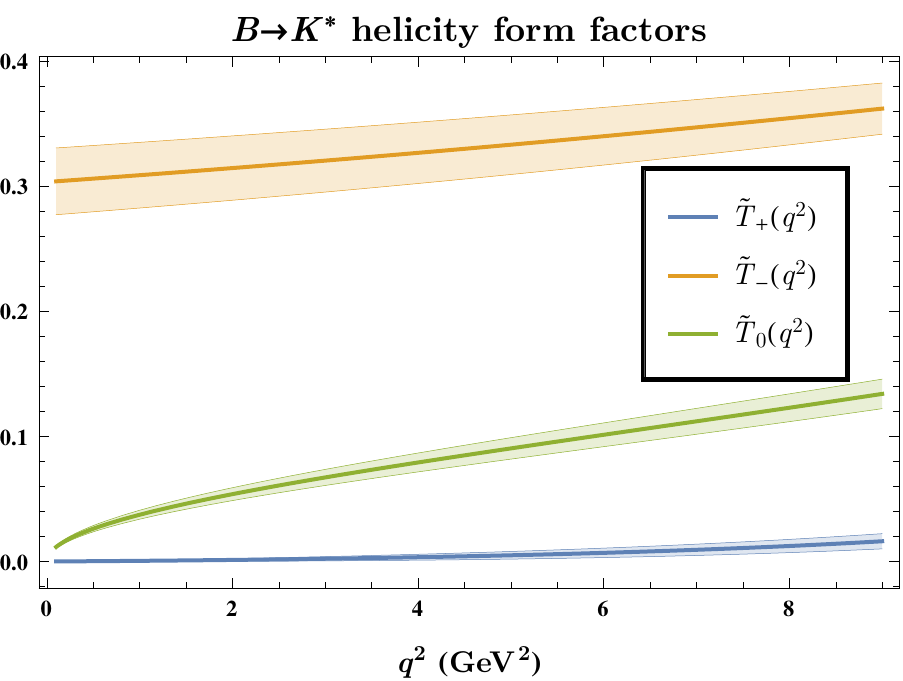}\vspace*{.4cm}
\caption{The central values and uncertainties of the $B\to K^*$ helicity form factors $\tilde{V}_{\pm,0}$ 
and $\tilde{T}_{\pm,0}$ using the LCSR results~\cite{Straub:2015ica}  
for the traditional form factors $V,A_{1,12}$ and $T_{1,2,23}$. 
\label{fig:VTtilde}}
\end{figure}

Considering the $q^2$-dependence of $h_\lambda(q^2)$, $\lambda=\pm 1,0$, it seems that $C_7$ ($C_9$) corresponds to the constant terms $h_\lambda^{(0)}$ ($h_\lambda^{(1)}$). 
However, such a one-to-one correspondence is not possible since the helicity form factors    
multiplying $C_7$ and $C_9$ in $H_V(\lambda)$ have a $q^2$-dependence themselves as 
can be seen in Fig.~\ref{fig:VTtilde}, in particular $\tilde{T}_0$ and $\tilde{V}_0$ have a significant $q^2$ dependence.  
Moreover, even when assuming the helicity form factors being trivial (means constant, independent of $q^2$), the interpretation of the three coefficients $h_\lambda^{(0)}$ ($h_\lambda^{(1)}$) as NP contribution to $C_7$  ($C_9$) is only possible if all these three coefficients are equal.  

The helicity form factors can be well described as 
\begin{align}\label{eq:3termexpansion}
 \tilde{T}_\lambda&= \frac{m_B}{2\hat{m}_b} (-16\pi^2) \left( a^{\tilde{T}}_\lambda +b^{\tilde{T}}_\lambda q^2 +c^{\tilde{T}}_\lambda q^4 \right)\,,\nonumber \\ 
 \tilde{V}_\lambda&= m_B^2 \frac{-16\pi^2}{q^2} \left( a^{\tilde{V}}_\lambda +b^{\tilde{V}}_\lambda q^2 +c^{\tilde{V}}_\lambda q^4 \right)\,,
\end{align}
where $a_\lambda^{\tilde{T},\tilde{V}},b_\lambda^{\tilde{T},\tilde{V}},c_\lambda^{\tilde{T},\tilde{V}}$ are fixed numbers that are obtained by expanding the form factors $\tilde{T}_{\lambda}$ and $\tilde{V}_{\lambda}$.
It should be noted that a simpler $q^2$-expansion without the  $c_\lambda^{\tilde{T},\tilde{V}}$ terms does not reasonably describe the form factors. 
Using the expansion of Eq.~(\ref{eq:3termexpansion}), the NP effect on the helicity amplitudes (via contributions to  $C_7$ or $C_9$)   
can be written as
\begin{align}\label{eq:deltaHVNP}
\delta H_V^{\lambda,{C_7^{\rm NP}}} &=  -i N^\prime \frac{2\hat{m}_b m_B}{q^2} 
\tilde{T}_{\lambda}(q^2)
C_7^{\rm NP} = i N^\prime m_B^2 \frac{16 \pi^2}{q^2} 
\left(a^{\tilde{T}}_\lambda C_7^{\rm NP}  +
q^2 b^{\tilde{T}}_\lambda C_7^{\rm NP}   + 
q^4 c^{\tilde{T}}_\lambda C_7^{\rm NP} \right)\,,\nonumber \\
\delta H_V^{\lambda,{C_9^{\rm NP}}} &=  -i N^\prime \tilde{V}_{\lambda}(q^2)
C_9^{\rm NP} = i N^\prime m_B^2 \frac{16 \pi^2}{q^2} 
\left(a^{\tilde{V}}_\lambda C_9^{\rm NP}  +
q^2 b^{\tilde{V}}_\lambda C_9^{\rm NP}   + 
q^4 c^{\tilde{V}}_\lambda C_9^{\rm NP} \right)\,.
\end{align}

Assuming the NP contributions to $C_7$ and $C_9$ to be complex numbers, there are overall four free parameters involved.
Comparing Eq.~(\ref{eq:deltaHVPC}) with Eq.~(\ref{eq:deltaHVNP}), it is clear that the NP effect can be embedded in the more general case of the hadronic effects.
Therefore, any NP effect could be simulated by some hadronic effect.
However, clearly not every hadronic effect can be described via some NP contribution.
But while it might be considered that any NP fit is just a specific case of the hadronic effects, this would be 
unlikely since hadronic power corrections have no reason to appear in the three helicity amplitudes $H_V^{\pm,0}$, 
in the same way as a single $C_9^{\rm NP}$ (and/or $C_7^{\rm NP}$) contribution, and it would be peculiar if 
they all conspired to have such an effect. Thus, a direct statistical comparison between a fit to NP and to power corrections which we present in the next section can be very illuminating.

In this context the criterion introduced in Ref.~\cite{Ciuchini:2015qxb} should be reanalysed. The authors claim that a large $q^2$ dependence, in particular the need of $q^4$ terms in the hadronic fit ansatz, disfavours or even disproves the NP interpretation of the LHCb anomalies. However, in view of the discussion above, we emphasise that a $q^4$ term could also be produced by a NP contribution, in particular in the helicity amplitude $H_V^0$ due to the $q^2$ dependence of the helicity form factors. Thus, such a criterion as proposed in Ref.~\cite{Ciuchini:2015qxb} can still be established, but has just to be tightened: If one finds a $q^2$ dependence in the hadronic fit which cannot be produced by any NP contribution then it is still possible to disfavour or even disprove the NP interpretation. 

On the other hand, a modest $q^2$ dependence should not be misinterpreted as strong indication for a NP resolution of the LHCb anomalies, because for example possible resonances might be smeared out via the large binning (2 GeV$^2$) of the experimental data. 
And in general it is true that as long as the NP fit is embedded in the more  general hadronic fit, as it is constructed now, one cannot disprove the latter option in favour of the former one with the present set of observables. 

%% file: results.tex
\section{Results}\label{sec:results}
In this analysis we consider the LHCb dataset on $B \to K^* \mu^+ \mu^-$ 
including the angular\footnote{
For the angular observables we consider
the LHCb results determined by the maximum likelihood fit method.}
observables~\cite{Aaij:2015oid} and the branching
ratios~\cite{Aaij:2016flj} using the low-$q^2$ bins up to either $q^2$= 6 or 8 GeV$^2$, which results in 36 or 45 observables, respectively.
The QCDf approach is only viable in the low $q^2$ region, where the calculations are most reliable for $q^2 \lesssim 7$ GeV$^2$~\cite{Beneke:2001at}. Thus, we have given our results for the two cases were the experimental data of the $[6,8]$ GeV$^2$ bins are included or disregarded.
The theoretical predictions are obtained using SuperIso v3.6~\cite{Mahmoudi:2007vz,Mahmoudi:2008tp}.

In order to determine whether underestimated hadronic effects or new physics contributions to $C_9$  better explain the observed tensions between the experimental results and SM predictions,
we compare a NP fit to $C_9$ with a fit to hadronic power corrections.
We perform the fits minimising the $\chi^{2}$ function provided by SuperIso, using the MINUIT minimisation tool~\cite{Lazzaro:2010zza}. 

%% file: results_details.tex
\subsection{Details on the fits}\label{subsec:details}
\subsubsection{Fit results for $C_9$}

\begin{figure} [h!]
\begin{center}
\includegraphics[scale=0.30]{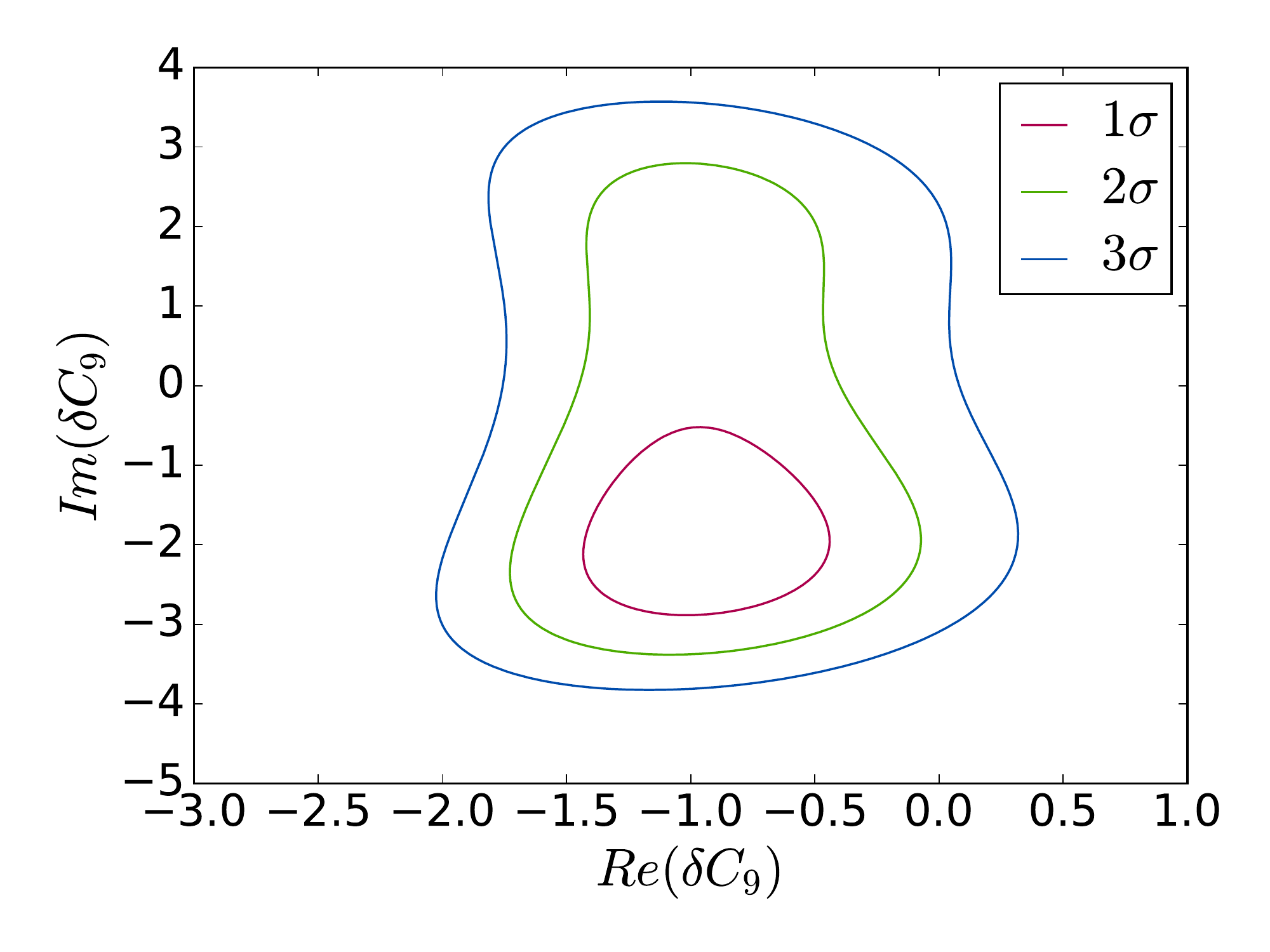} 
\includegraphics[scale=0.30]{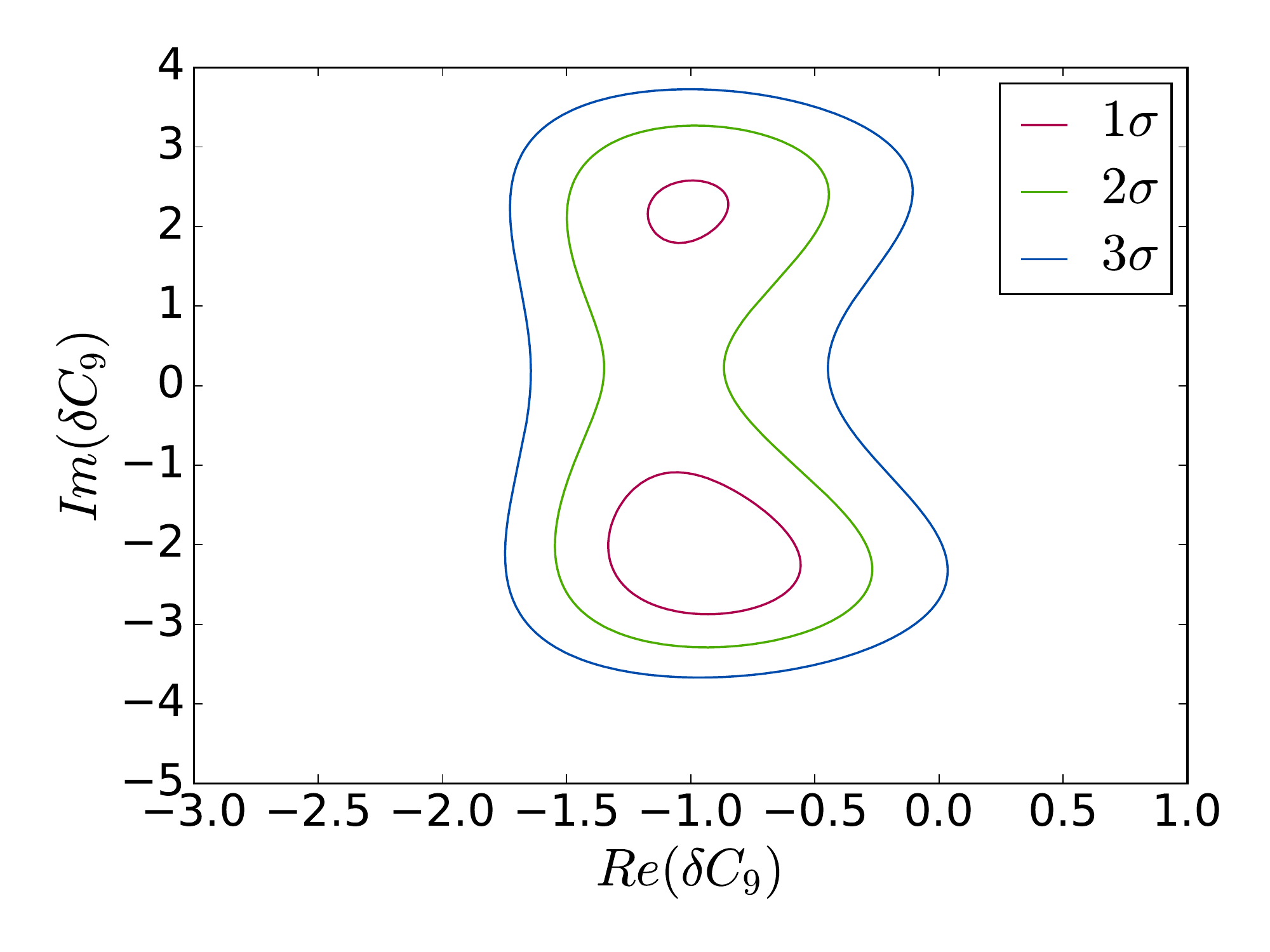} 

\caption{Contour plots for real and imaginary parts of $\delta C_9$.
The contours on the left correspond to fits up to $q^2 = 6$ GeV$^2$, and the contours on the right correspond to fits up to $q^2=8$ GeV$^2$.
 \label{fig:contC9}}
\end{center}
\end{figure}

We first perform a new physics fit in which only the Wilson coefficient $C_9^{\rm NP}$ is allowed to differ from zero. 
The best fit point for a complex $C_9^{\rm NP}$ is given in Table~\ref{tab:c9} for the two cases, in which bins up to $q^2=6$ GeV$^2$ or up to $q^2=8$ GeV$^2$ are considered.
The best fit value for $C_9$ remains almost the same for both cases.
This is an interesting result, given that the theory predictions are less reliable for the $[6,8]$ GeV$^2$ bins compared to the region below 6 GeV$^2$, and also given that one of the deviations of $P_5^\prime$ is in the $[6,8]$ GeV$^2$ bin.
The best fit value of $C_9$ is consistent with the case when all the $b \to s \ell^+ \ell^-$ data is used. 
The two-dimensional contours for $\delta C_9$ are given in Fig.~\ref{fig:contC9}.
The contours for the cases up to $q^2 = 6$ GeV$^2$ and $q^2=8$ GeV$^2$ have similar shapes, 
with the latter being slightly narrower for the real parts. 
The main difference is in $Im(C_9)$ where two distinct $1\sigma$ regions are possible for the fit using up to $q^2=8$ GeV$^2$. 

\begin{table}[t!]
\ra{1.}
\rb{1.3mm}
\begin{center}
\setlength\extrarowheight{2pt}
\begin{tabular}{|c|c|}
\hline
 & \multicolumn{1}{c|}{\footnotesize{up to $q^2=6$ GeV$^2$ obs.}}           \\  
\hline
$Re(\delta C_9)$ & $-0.96_{-0.32}^{+0.34}$   \\ 
$Im(\delta C_9)$ & $-1.96_{-0.64}^{+0.82}$ \\                                                                                        
\hline
\end{tabular} 
\begin{tabular}{|c|c|}
\hline
 & \multicolumn{1}{c|}{\footnotesize{up to $q^2=8$ GeV$^2$ obs.}}           \\  
\hline
$Re(\delta C_9)$ &  $-0.97 _{-0.25}^{+0.26}$  \\
$Im(\delta C_9)$ &  $-2.13_{-0.50}^{+0.62}$ \\                                                                                        
\hline
\end{tabular} 
\caption{Fit results for $\delta C_9$ alone using observables up to $q^2 = 6$ GeV$^2$ as well as up to $q^2 = 8$ GeV$^2$.
\label{tab:c9}
}
\end{center} 
\end{table}  

\subsubsection{Fit results for $\{C_7-C_9\}$}
\begin{figure} [htb!]
\begin{center}
\includegraphics[scale=0.30]{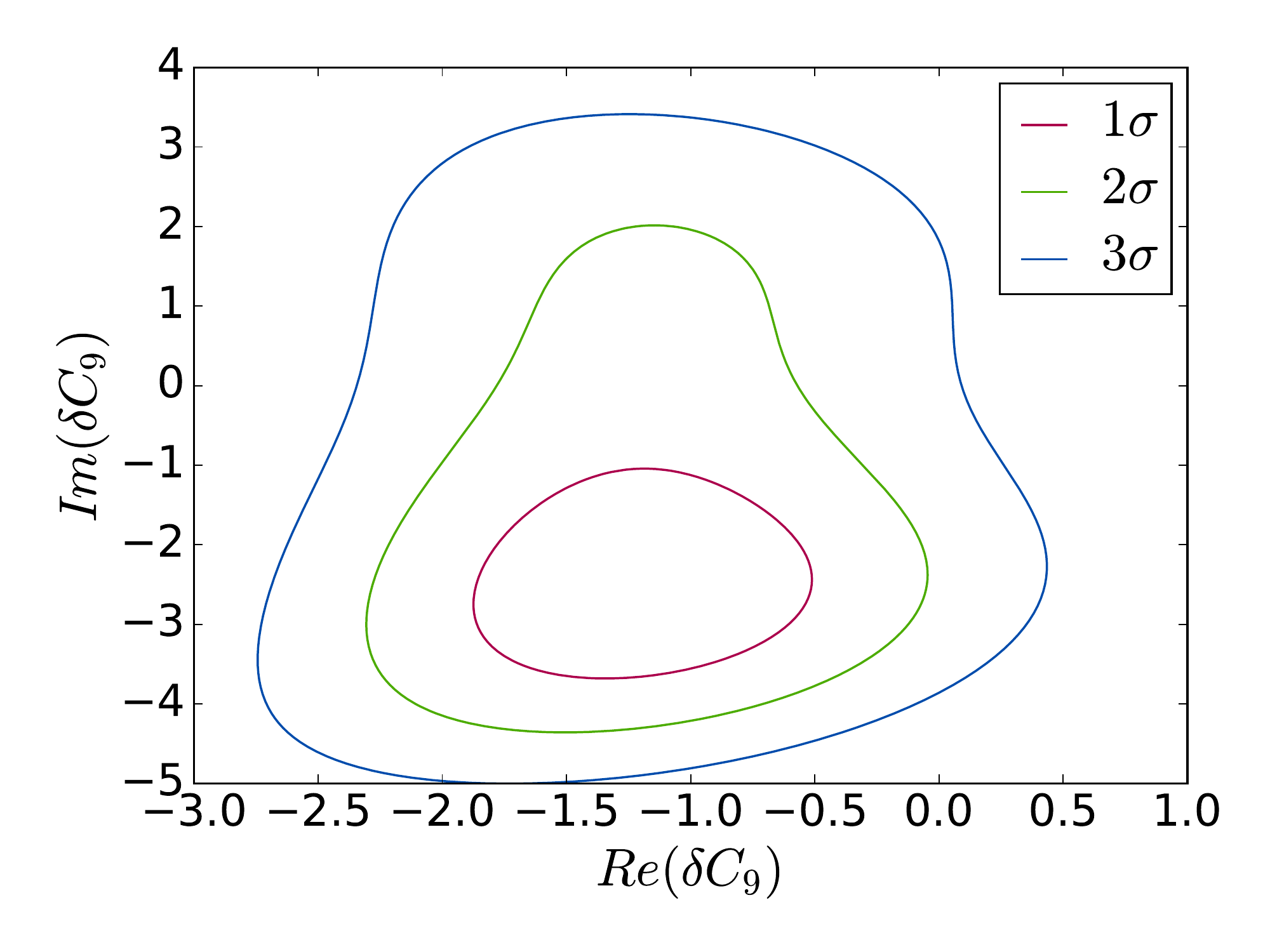} 
\includegraphics[scale=0.30]{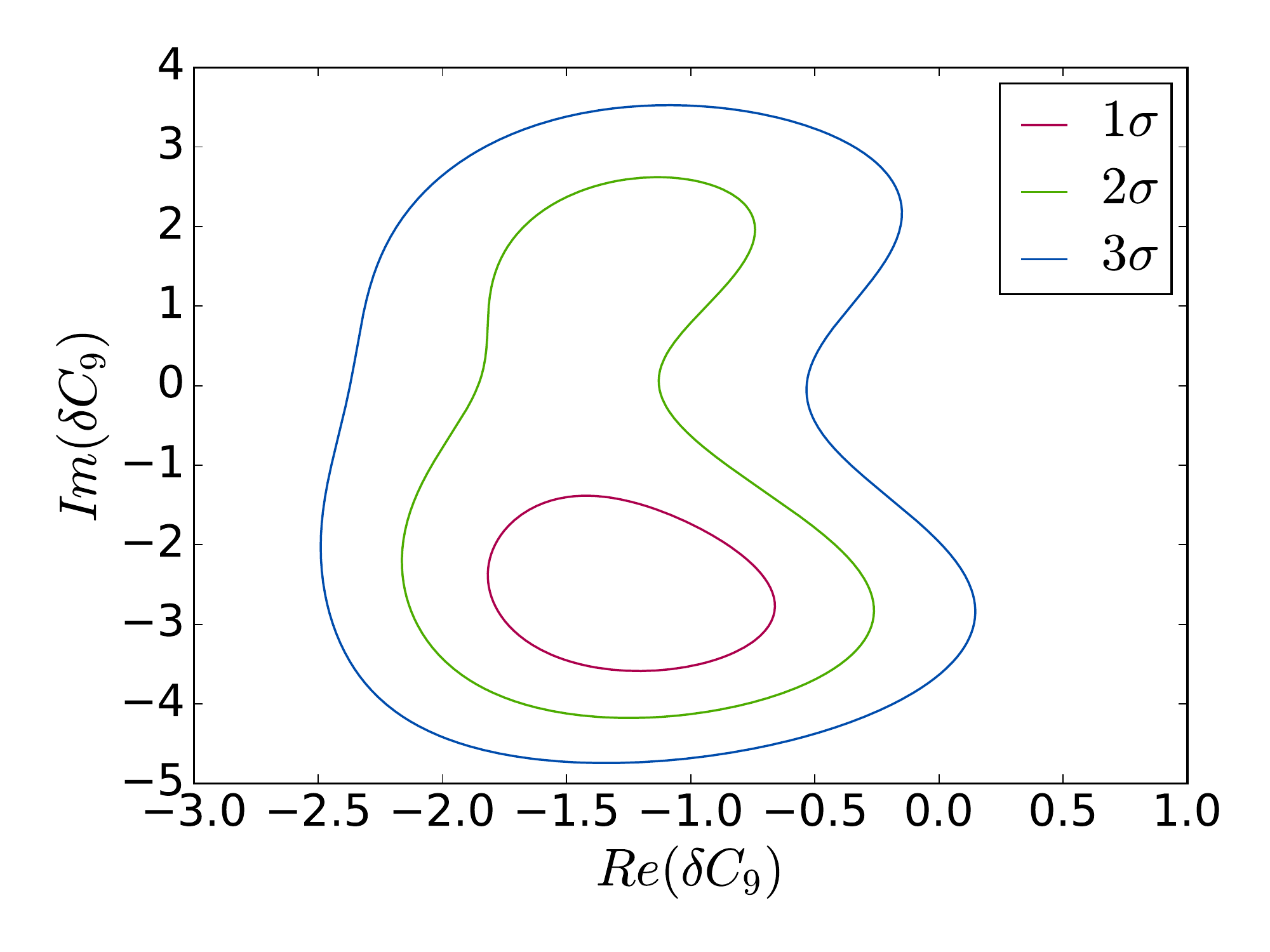} 

\caption{Contour plots for real and imaginary parts of $\delta C_9$ when both $\delta C_9$ and $\delta C_7$ are allowed to differ from zero.
For each plane, the $\chi^2$ is minimised over the remaining two parameters.
The contours on the left correspond to fits up to $q^2 = 6$ GeV$^2$, and the contours on the right correspond to fits up to $q^2=8$ GeV$^2$.
\label{fig:cont4}}
\end{center}
\end{figure}

We now consider the case where both $\delta C_7$ and $\delta C_9$ are allowed to vary simultaneously.
The values of the fitted parameters are shown in Table~\ref{tab:c7_6&8}, 
where the errors correspond to $\Delta\chi^2=1$ profiling over the remaining parameters. 
The best fit points are consistent with the fit in which only $C_9$ is allowed to vary.
The linear correlation between $Im(\delta C_9)$ and $Re(\delta C_9)$ is very small. 
In this case, $Im(\delta C_9)$ has only one $1\sigma$ region for the $q^2 = 8$ GeV$^2$ fit. 
This can be seen in the two-dimensional contours shown in Fig.~\ref{fig:cont4} for the two different fit configurations.

The constraint on $C_7$ obtained by considering only the $B\to K^*\mu^+\mu^-$ data (the current study) is significantly weaker 
than the constraint induced from the $b \to s \gamma$ data. 
However, we checked that the crucial constraint on $C_9$ in the $C_7-C_9$ fits does not change much between the two sets of constraints on $C_7$.

\begin{table}[t!]
\ra{1.2}
\rb{1.3mm}
\begin{center}
\setlength\extrarowheight{2pt}
\scalebox{0.74}{
\begin{tabular}{|c|c|c|c|c|c|}
\hline
 \multicolumn{6}{|c|}{up to $q^2=6$ GeV$^2$ obs.}           \\  
\hline
&value & $Re(\delta C_7)$ &$Im(\delta C_7)$ & $Re(\delta C_9)$ & $Im(\delta C_9)$ \\ 
\hline
$Re(\delta C_7)$ & $0.019_{-0.038}^{+0.039}$ & 1 & 0.06 & -0.68 & 0.05 \\
$Im(\delta C_7)$ &  $0.085_{-0.054}^{+0.056}$ & 0.06& 1 & -0.18 & -0.54 \\                                                               
$Re(\delta C_9)$ &$-1.20_{-0.44}^{+0.45}$  & -0.68 & -0.18 & 1 & 0.11 \\
$Im(\delta C_9)$ & $-2.54_{-0.77}^{+0.90}$ & 0.05 & -0.54 & 0.11 & 1 \\                                                                                        
\hline
\end{tabular} 
\begin{tabular}{|c|c|c|c|c|c|}
\hline
 \multicolumn{6}{|c|}{up to $q^2=8$ GeV$^2$ obs.}           \\  
\hline
&value & $Re(\delta C_7)$ &$Im(\delta C_7)$ & $Re(\delta C_9)$ & $Im(\delta C_9)$ \\ 
\hline
$Re(\delta C_7)$ & $0.028_{-0.037}^{+0.039}$ & 1 & 0.02 & -0.76 & 0.22 \\
$Im(\delta C_7)$ &  $0.097_{-0.051}^{+0.053}$ & 0.02& 1 & -0.09 & -0.55 \\                                                               
$Re(\delta C_9)$ &$-1.25_{-0.38}^{+0.39}$  & -0.76 & -0.09 & 1 & -0.19 \\
$Im(\delta C_9)$ & $-2.61_{-0.65}^{+0.75}$ & 0.22 & -0.55 & -0.19 & 1 \\                                                                                        
\hline
\end{tabular} 
}
\caption{Fit results and correlation coefficients for $\delta C_7$ and $\delta C_9$ using observables up to $q^2 = 6$ GeV$^2$ in 
the left table and up to $q^2 = 8$ GeV$^2$ observables in the right table. 
\label{tab:c7_6&8}
}
\end{center} 
\end{table}  
%

\subsubsection{Fit results for hadronic power corrections}
The results of the power correction fit for $h_{\pm,0}^{(0,1,2)}$ using the experimental $q^{2}$ bins up to $6$ GeV$^2$ and also up to $8$ GeV$^2$ bins are given in Table~\ref{tab:hParamsFree}\footnote{Our results are somewhat different compared to Table 5 of Ref.~\cite{Ciuchini:2015qxb};
we are using the frequentist approach and fitting the $h_\lambda^{(0,1,2)}$ parameters in Cartesian coordinates while Ref.~\cite{Ciuchini:2015qxb} follows the Bayesian approach and fitting in polar coordinates.
Moreover, we have not considered any constraint on $h_\lambda^{(0,1,2)}$ as opposed to Ref.~\cite{Ciuchini:2015qxb} (for our fit results where we consider the $|h_+^{(0)}/h_-^{(0)}|<0.2$, see appendix~\ref{app:constrained}).
There are also slight theoretical differences, where we have employed the updated LCSR form factor calculations of Ref.~\cite{Straub:2015ica}.
In addition, for the experimental results on BR$(B\to K^* \mu^+ \mu^-)$ we have used the updated LHCb measurement~\cite{Aaij:2016flj} and we do not include BR$(B\to K^* \gamma)$ which has been considered by Ref.~\cite{Ciuchini:2015qxb}.
Given these  differences, the magnitude and sign of the real values of our  hadronic fit are in good agreement with the results of Ref.~\cite{Ciuchini:2015qxb}.
However, the imaginary values have in many cases opposite signs with somewhat more differing magnitudes.}. 
No constraint is assumed for either of the 18 parameters. In this case, since many parameters are consistent with zero, we fit the real and imaginary parts ({\it i.e.}, in Cartesian coordinates) instead of magnitude and phase ({\it i.e.}, polar coordinates) which leads to serious convergency issues for magnitudes consistent with zero.

\begin{figure}[h!]
\centering
\includegraphics[width=0.7\textwidth]{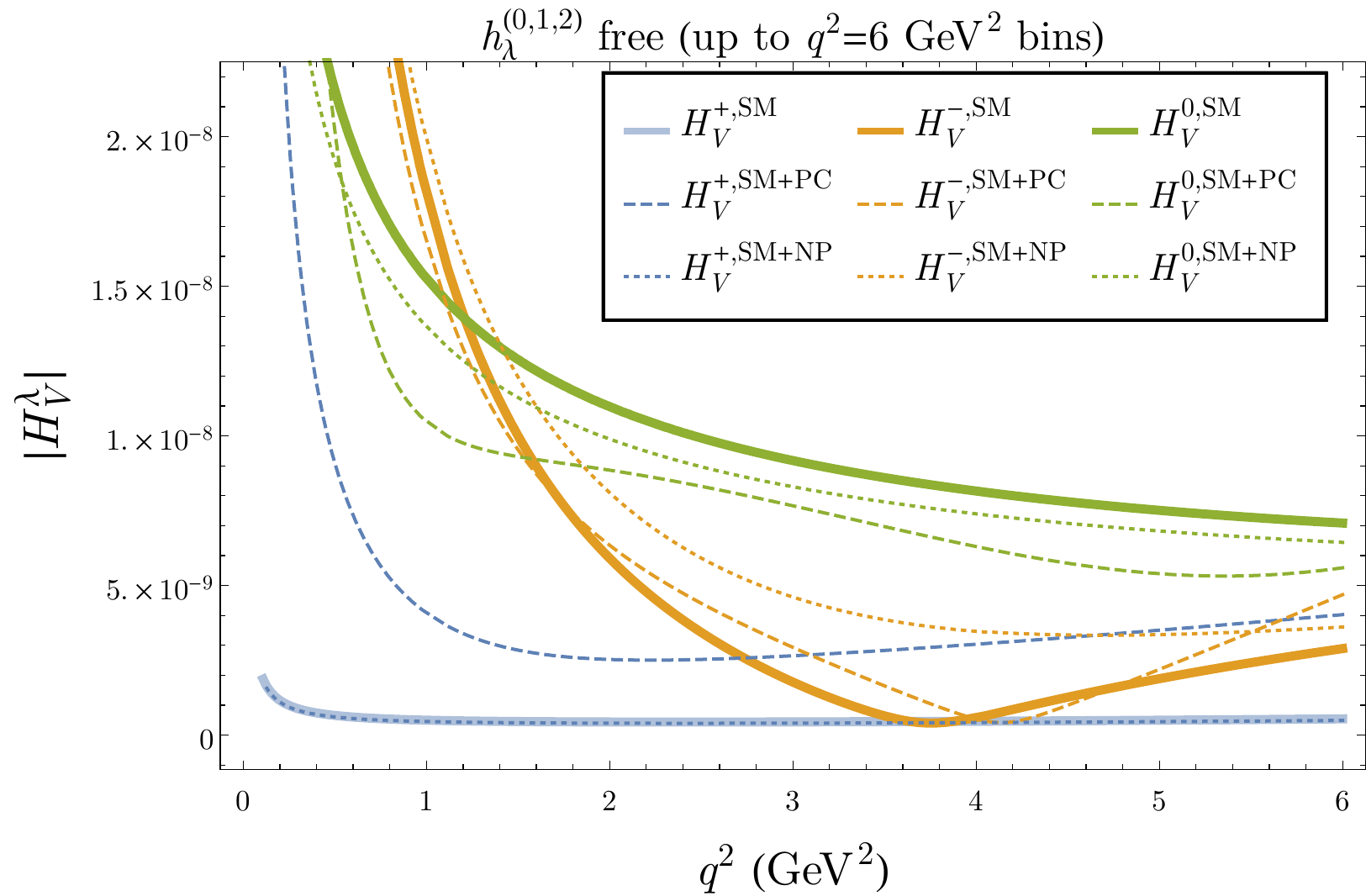}
\includegraphics[width=0.7\textwidth]{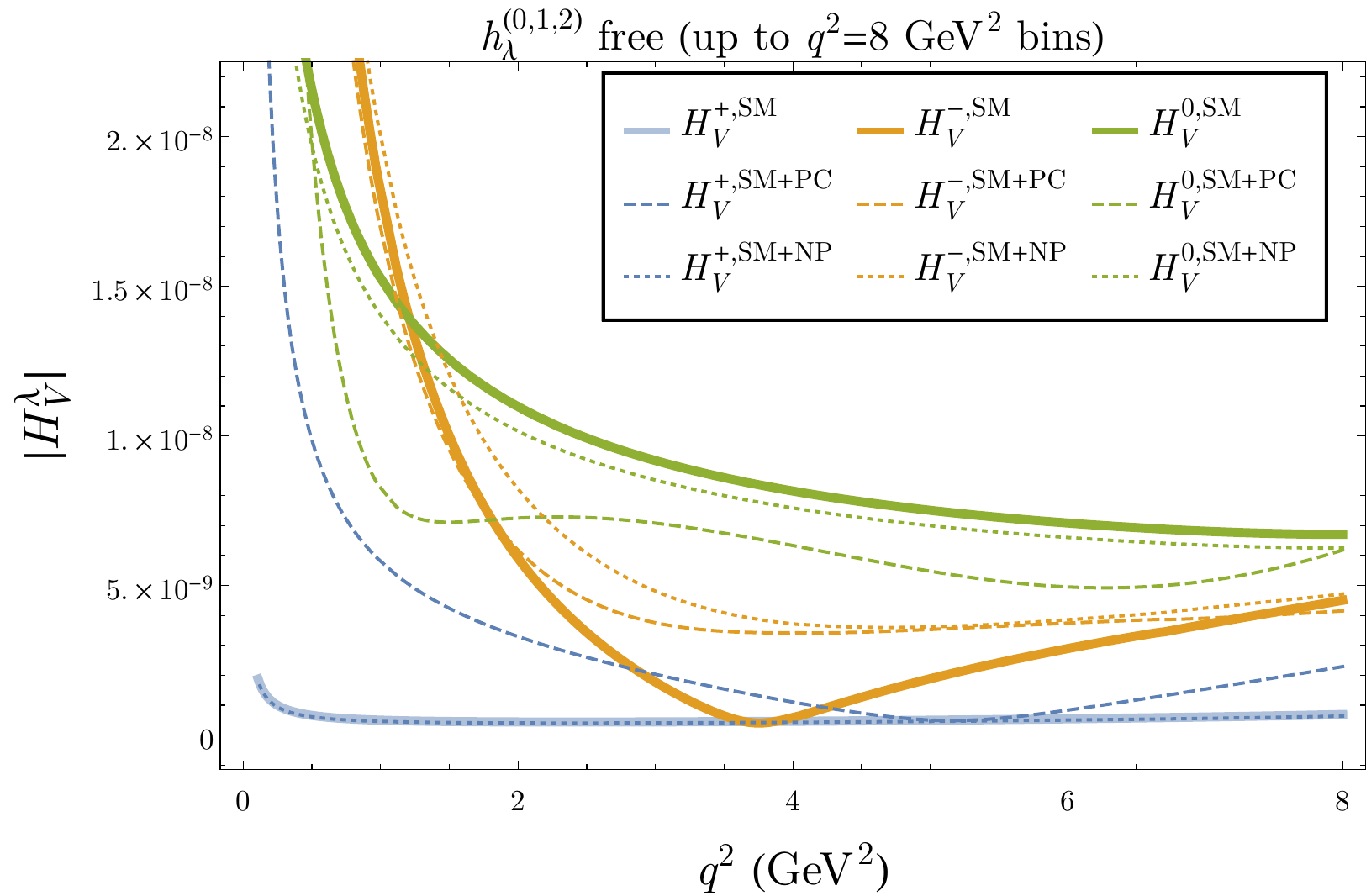}
\caption{Behaviour of absolute value of \textcolor{myblue}{$H_V^+$}, \textcolor{myorange}{$H_V^-$} and \textcolor{mygreen}{$H_V^0$}.
The ``SM'',  ``SM + power correction'' and ``SM+NP'' are shown with solid, dashed and dotted lines, respectively.
No constraint is assumed for $h_{\lambda}^{(0,1,2)}$ in the fit for power corrections.
\label{fig:HVallAbsFree}}
\end{figure}

The statistical comparison of the fit to  NP and to power corrections is given in the following subsections.
In Fig.~\ref{fig:HVallAbsFree}, the central values of $|H_V^{\pm,0}(q^2)|$ when including
the effect of the new physics and also the power corrections are compared with 
the plain SM predictions.
As discussed in the previous section, the $q^2$-dependence that $h_\lambda^{(2)}$ introduces can
potentially also be produced by NP contributions to $C_9$ due to the form factors. 
However, in case the $q^2$-dependence of the power corrections for $H_V$ is significantly different
compared to the corresponding NP fit then they could be differentiable. 
Nonetheless, in Fig.~\ref{fig:HVallAbsFree},
the shapes for both the effects of NP and hadronic power corrections in $H_V$ are rather similar and the $q^2$ shape
is only significantly different for very low $q^2$.

\begin{table}[t!]
\ra{0.90}
\rb{1.3mm}
\begin{center}
\setlength\extrarowheight{2pt}
\footnotesize{
\begin{tabular}{|l||c|c|}
\hline
 \multicolumn{3}{|c|}{up to $q^2=6$ GeV$^2$ obs.}           \\  
\hline
& Real                          & Imaginary    \\ 
 \hline
$h_+^{(0)}$     & $ ( 2.3 \pm   2.3) \times 10^{-4}$   &  $( -2.0 \pm   2.3) \times 10^{-4}$ \\ 
$h_+^{(1)}$     & $ ( -1.2 \pm  3.5) \times 10^{-4}$   &  $( 3.3 \pm  38.6) \times 10^{-5}$ \\ 
$h_+^{(2)}$     & $ ( 1.2 \pm  6.8) \times 10^{-5}$    &  $( -3.5 \pm  8.1) \times 10^{-5}$   \\ 
\hline 
$h_-^{(0)}$     & $ ( -7.7 \pm  19.8) \times 10^{-5}$  &  $( 4.5 \pm  3.6) \times 10^{-4}$  \\ 
$h_-^{(1)}$     & $ ( -3.7 \pm  20.8) \times 10^{-5}$  &  $( -7.4 \pm  4.2) \times 10^{-4}$   \\ 
$h_-^{(2)}$     & $ ( 2.7 \pm  3.9) \times 10^{-5}$    &  $( 1.5 \pm  0.8) \times 10^{-4}$  \\ 
\hline
$h_0^{(0)}$     & $ ( -6.1 \pm  38.4) \times 10^{-5}$  &  $( 7.8 \pm  4.0) \times 10^{-4}$  \\ 
$h_0^{(1)}$     & $ ( 3.8 \pm   5.2) \times 10^{-4}$   &  $( -1.0 \pm  0.6) \times 10^{-3}$   \\ 
$h_0^{(2)}$     & $ ( -4.7 \pm  8.7) \times 10^{-5}$   &  $( 1.6 \pm  1.3) \times 10^{-4}$  \\ 
\hline
\end{tabular} 
}
\footnotesize{
\begin{tabular}{|l||c|c|}
\hline
 \multicolumn{3}{|c|}{up to $q^2=8$ GeV$^2$ obs.}           \\  
\hline
& Real                          & Imaginary    \\ 
 \hline
$h_+^{(0)}$     & $ (1.2 \pm  2.0) \times 10^{-4}$   &  $(-1.6 \pm  2.1) \times 10^{-4}$ \\ 
$h_+^{(1)}$     & $ (1.2 \pm  2.3) \times 10^{-4}$   &  $(-1.1 \pm  3.0) \times 10^{-4}$ \\ 
$h_+^{(2)}$     & $ (-2.6 \pm  3.4) \times 10^{-5}$  &  $(2.3 \pm  4.4) \times 10^{-5}$   \\ 
\hline 
$h_-^{(0)}$     & $ (-1.0 \pm  1.8) \times 10^{-4}$  &  $(2.9 \pm  3.2) \times 10^{-4}$  \\ 
$h_-^{(1)}$     & $ (2.5 \pm  13.3) \times 10^{-5}$  &  $(-3.4 \pm  3.2) \times 10^{-4}$   \\ 
$h_-^{(2)}$     & $ (9.2 \pm  18.7) \times 10^{-6}$  &  $(1.7 \pm  4.8) \times 10^{-5}$  \\ 
\hline
$h_0^{(0)}$     & $ (-2.6 \pm  3.3) \times 10^{-4}$  &  $(6.5 \pm  3.9) \times 10^{-4}$  \\ 
$h_0^{(1)}$     & $ (7.5 \pm  4.4) \times 10^{-4}$   &  $(-8.7 \pm  3.6) \times 10^{-4}$   \\ 
$h_0^{(2)}$     & $ (-8.6 \pm  5.8) \times 10^{-5}$  &  $(9.6 \pm  6.2) \times 10^{-5}$  \\ 
\hline
\end{tabular} 
}
\caption{Fit parameters for the power corrections  assuming no constraint for $h_{\lambda}^{(0,1,2)}$. 
\label{tab:hParamsFree}
}
\end{center} 
\end{table}  

\clearpage

%% file: results_Wilktests1.tex
\subsection{Likelihood ratio tests for $\delta C_7$, $\delta C_9$ and hadronic fits }\label{subsec:Wilktest1}

The different models can be compared via likelihood ratio tests. Since they are nested models, $p$-values can be obtained via applying Wilks' theorem ~\cite{Wilks:1938dza}, 
where the difference in $\chi^2$ between the two models is itself a $\chi^2$ distribution with a number of degrees of freedom equal to the difference in number of parameters.
The $p$-value indicates therefore the significance of the new parameters added. 

Here, we directly compare the NP fit with the hadronic fit. The results obtained by this test are shown in Table~\ref{tab:wilk1_6&8}. For convenience, the $p$-values are translated to 
Gaussian single-parameter significances as
\begin{equation}
\sigma = \sqrt{2}Erf^{-1}(1-p)\;.
\end{equation}

\begin{table}[t!]
\ra{1.2}
\rb{1.3mm}
\begin{center}
\setlength\extrarowheight{2pt}
\scalebox{0.73}{
\begin{tabular}{|c|c|c|c|}
\hline
 \multicolumn{4}{|c|}{up to $q^2=6$ GeV$^2$ obs.}           \\  
\hline
& $\delta C_9$ & $\delta C_7,\delta C_9$ & Hadronic\\
\hline
plain SM & $4.5\times 10^{-3}(2.8\sigma)$ & $9.4\times10^{-3}(2.6\sigma)$ & $6.2\times 10^{-2}(1.9\sigma)$\\
$\delta C_9$ & -- & $0.27 (1.1\sigma)$ & $0.37(0.89\sigma)$\\
$\delta C_7,\delta C_9$ & -- & -- & $0.41(0.86\sigma)$ \\
\hline
\end{tabular} 
}
\scalebox{0.73}{
\begin{tabular}{|c|c|c|c|c|c|}
\hline
 \multicolumn{4}{|c|}{up to $q^2=8$ GeV$^2$ obs.}           \\  
\hline
& $\delta C_9$ & $\delta C_7,\delta C_9$ & Hadronic\\
\hline
plain SM & $3.7\times 10^{-5}(4.1\sigma)$ & $6.3\times10^{-5}(4.0\sigma)$ & $6.1\times 10^{-3}(2.7\sigma)$\\
$\delta C_9$ & -- & $0.13 (1.5\sigma)$ & $0.45(0.76\sigma)$\\
$\delta C_7,\delta C_9$ & -- & -- & $0.61(0.52\sigma)$ \\
\hline
\end{tabular} 
}
\caption{Likelihood ratio $p$-values and significances obtained using Wilks' theorem for observables up to $q^2 = 6$ GeV$^2$ and $q^2 = 8$ GeV$^2$ in the left and right table, respectively.
\label{tab:wilk1_6&8}
}
\end{center} 
\end{table}  

\begin{table}[t!]
\ra{1.2}
\rb{1.3mm}
\begin{center}
\setlength\extrarowheight{2pt}
\scalebox{0.73}{
\begin{tabular}{|c|c|c|}
\hline
 \multicolumn{3}{|c|}{up to $q^2=6$ GeV$^2$ obs.}           \\  
\hline
& $\delta C_9$ & $\delta C_7,\delta C_9$ \\
\hline
plain SM & $5.3\times 10^{-3}(2.8\sigma)$ & $5.3\times10^{-3}(2.8\sigma)$ \\
\hline
\end{tabular} 
}
\scalebox{0.73}{
\begin{tabular}{|c|c|c|}
\hline
 \multicolumn{3}{|c|}{up to $q^2=8$ GeV$^2$ obs.}           \\  
\hline
& $\delta C_9$ & $\delta C_7,\delta C_9$\\
\hline
plain SM & $3.4\times 10^{-5}(4.1\sigma)$ & $5.5\times10^{-5}(4.0\sigma)$\\
\hline
\end{tabular} 
}
\caption{Likelihood ratio $p$-values and significances obtained through likelihood integration using observables up to $q^2 = 6$ GeV$^2$ and $q^2 = 6$ GeV$^2$ in the left and right table, respectively.
\label{tab:wilk1_6&8_1}
}
\end{center} 
\end{table}  

Adding $\delta C_9$ improves over the SM hypothesis by $2.8\sigma$ ($4.1\sigma$) for $q^2 = 6$ GeV$^2$ ($q^2 = 8$ GeV$^2$). Including in addition $\delta C_7$ or 16 hadronic parameters improves the situation only mildly.
Therefore, the extra parameters of the more general models are not significant compared to a fit with only the complex parameter $\delta C_9$.

However, Wilks' theorem is only valid under the assumption of Gaussian distributed uncertainties. To verify whether it can be safely applied in our case, we calculate the $p$-values
in an independent way. In this test, the $p$-values of the plain SM versus $\delta C_9$ fit and versus the $\delta C_7,\delta C_9$ fit are obtained 
from likelihood integration from the best fit point down to the SM with the $\chi^2$ as ordering principle. We do not perform it on the hadronic power correction fit because 
it is computationally too expensive due to the large number of free parameters. The resulting significances are summarised in Tables~\ref{tab:wilk1_6&8_1}. They are very similar to those provided by
applying Wilks' theorem and hence we conclude that it is a reliable approach, which we directly apply on the hadronic power correction fit. 

At present the data can be well described by $C_9^{\rm NP}$ and including the general hadronic parameters does not improve the fit. 
The Wilks' tests suggest that the present data do not disfavour the NP option which is still a viable solution.

%% file: results_Wilktests2.tex
\subsection{Contours and likelihood ratio tests for $\delta C_9$ and $\delta C_{10}$ fits}\label{subsec:Wilktest2}

\begin{figure} [t!]
\begin{center}
\includegraphics[scale=0.30]{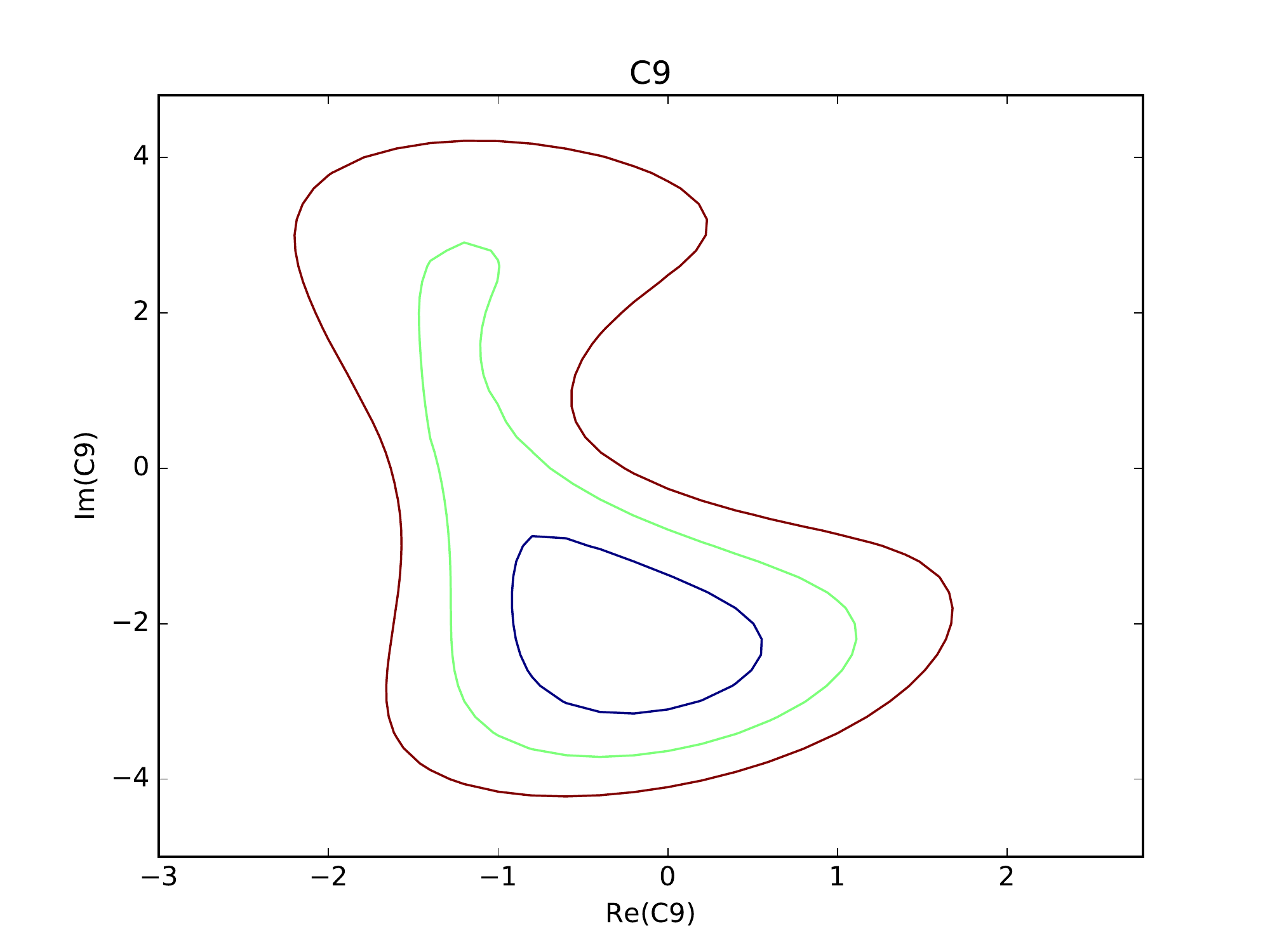} 
\includegraphics[scale=0.30]{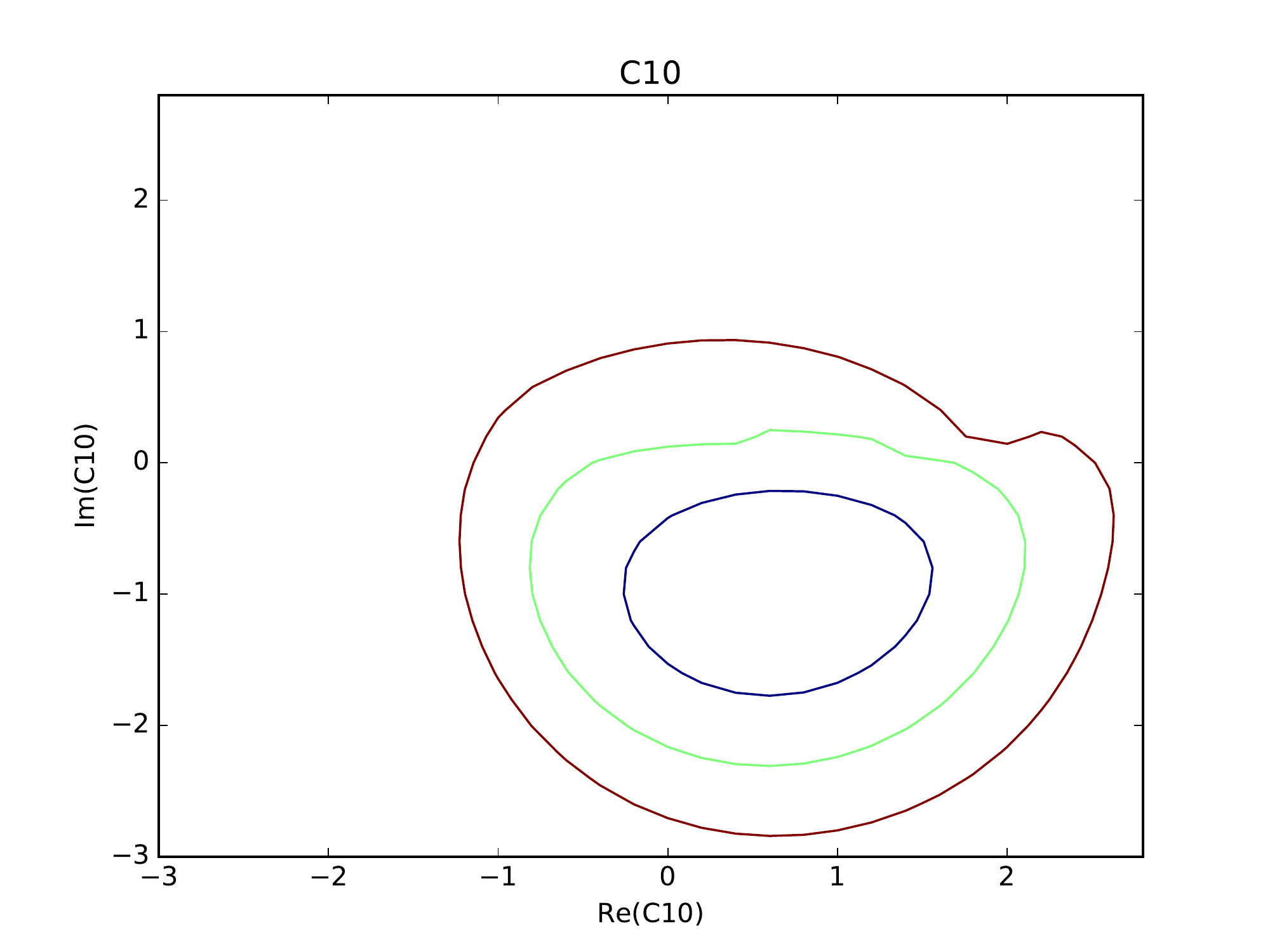} 
\caption{Contour plots for real and imaginary part of $\delta C_9$ (left) and $\delta C_{10}$ (right) when both coefficients are allowed to differ from zero. The $\chi^2$ at each point of the plane is minimised with respect to the remaining two variables.
 \label{fig:c10}}
\end{center}
\end{figure}

In this section we allow for BSM dynamics in the complex Wilson coefficient $\delta C_{10}$. This coefficient
can alter the decay rates of $B_{(s)}^{0}\rightarrow\mu^+\mu^-$ and thus we include the results
on this decay  
from ~\cite{CMS:2014xfa}
into our fits. 
Here, for the sake of conciseness, we show the results using up to $q^2=8$ GeV$^2$.
The values of the fitted parameters are shown in Table \ref{tab:c10}, where the errors correspond
to $\Delta\chi^2=1$ profiling over the remaining parameters.
\begin{table}[t!]
\ra{1.0}
\rb{1.3mm}
\begin{center}
\setlength\extrarowheight{2pt}
\footnotesize{
\begin{tabular}{|c|c|c|c|c|c|}
\hline
 \multicolumn{6}{|c|}{up to $q^2=8$ GeV$^2$ obs.}           \\  
\hline
&value & $Re(\delta C_9)$ &$Im(\delta C_9)$ & $Re(\delta C_{10})$ & $Im(\delta C_{10})$ \\ 
\hline
$Re(\delta C_9)$ & $-0.24_{-0.47}^{+0.51}$ & 1 & -0.27 & 0.65 & -0.36 \\
$Im(\delta C_9)$ &  $-2.19_{-0.66}^{+0.79}$ & -0.27& 1 & -0.57 & -0.60 \\                                                               
$Re(\delta C_{10})$ &$0.66_{-0.61}^{+0.59}$  & 0.67 & -0.57 & 1 & 0.08 \\
$Im(\delta C_{10})$ & $-0.97_{-0.52}^{+0.50}$ & -0.36 & -0.60 & 0.08 & 1 \\                                                                                        
\hline
\end{tabular} 
}
\caption{Fit results and correlation coefficients for $\delta C_9$ and $\delta C_{10}$ using observables up to $q^2 = 8$ GeV$^2$. 
\label{tab:c10}
}
\end{center} 
\end{table}  
%
The two-dimensional contours of $\delta C_9$ and $\delta C_{10}$ can be seen in Fig.~\ref{fig:c10}.

The $p$-values of the likelihood ratio tests are shown in Table \ref{tab:wilk2_8}. It can be seen that
there is no significant gain from using $\delta C_{10}$ in addition to $\delta C_9$. On the contrary, there is a significant gain
from using $\delta C_{10},\delta C_9$ with respect to $\delta C_{10}$ alone.
\begin{table}[t!]
\ra{1.0}
\rb{1.3mm}
\begin{center}
\setlength\extrarowheight{2pt}
\footnotesize{
\begin{tabular}{|c|c|c|}
\hline
 \multicolumn{3}{|c|}{up to $q^2=8$ GeV$^2$ obs.}           \\  
\hline
& $\delta C_9$ & $\delta C_9,\delta C_{10}$ \\
\hline
plain SM & $3.54\times10^{-5}(4.1\sigma)$ & $4.74\times10^{-5}(4.1\sigma)$ \\
$\delta C_9$ & -- & $0.099(1.7\sigma)$\\
$\delta C_{10}$&--& $4.31\times10^{-4}(3.5\sigma)$\\                                                
\hline
\end{tabular} 
}
\caption{Likelihood ratio $p$-values and significances obtained using Wilks' theorem for observables up to $q^2 = 8$ GeV$^{2}$.
\label{tab:wilk2_8}
}
\end{center} 
\end{table}  

%% file: prospects.tex
\subsection{LHCb upgrade prospects}\label{subsec:prospects}

\begin{figure} [t!]
\begin{center}
\includegraphics[scale=0.30]{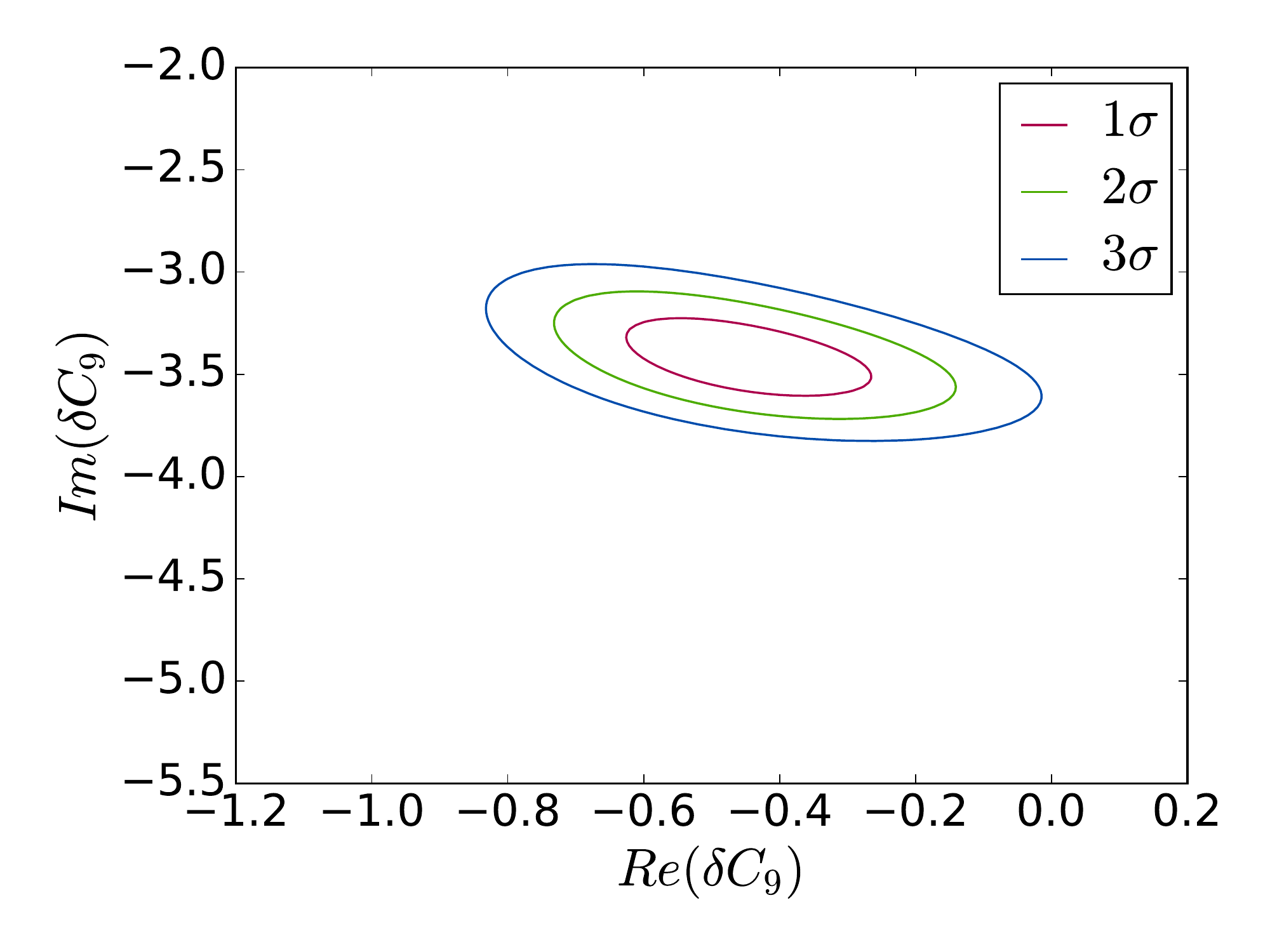} 
\includegraphics[scale=0.30]{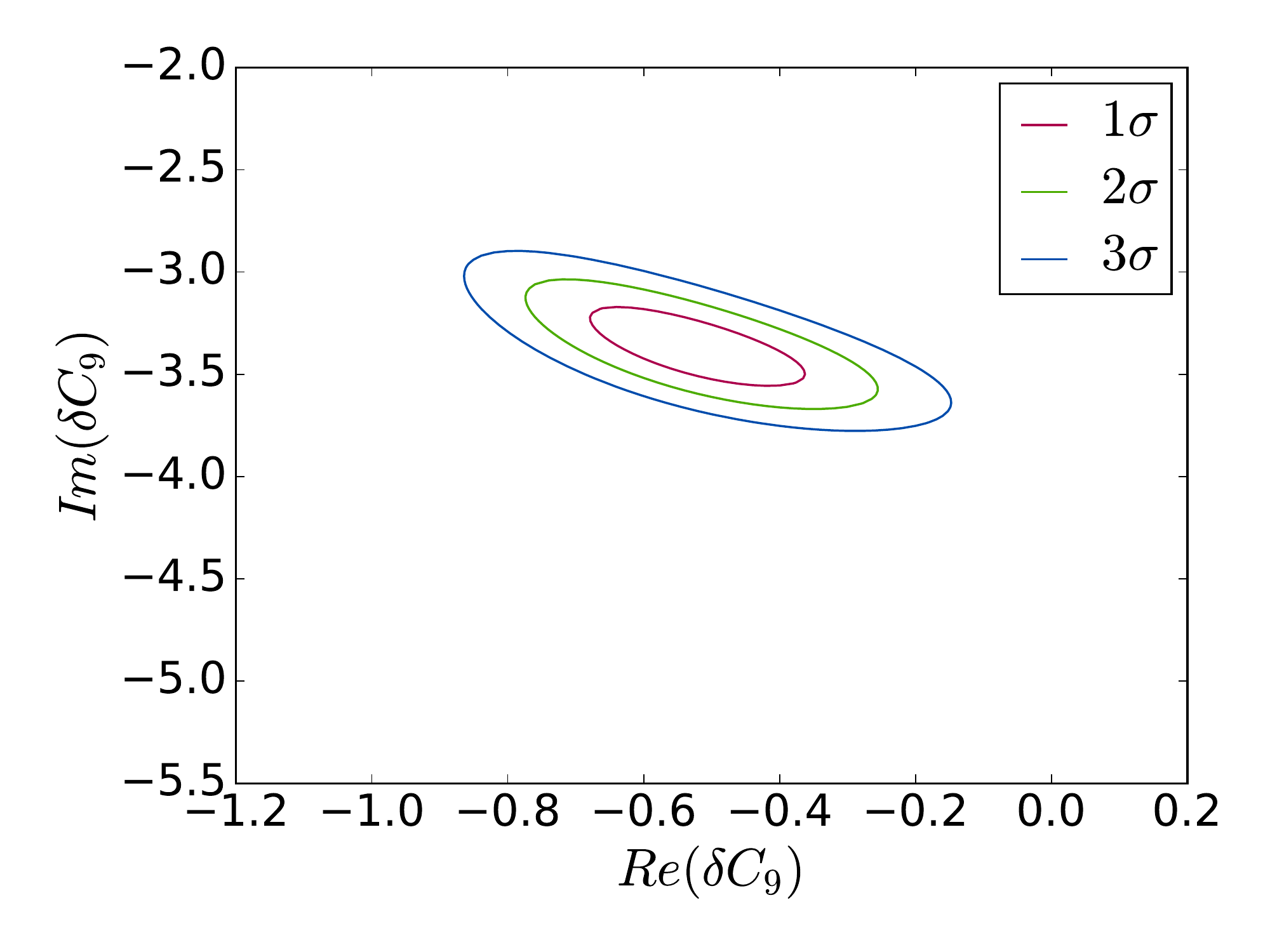} \\

\caption{Projections of the real and imaginary part of $\delta C_9$ assuming integrated luminosity of 300 fb$^{-1}$. Only $\delta C_9$ is allowed to differ from zero. The contours on the left correspond to fits up to $q^2 = 6$ GeV$^2$, and the contours on the right correspond to fits up to 
$q^2=8$ GeV$^2$. 
 \label{fig:upgrade2}}
\end{center}
\end{figure}

\begin{figure} [t!]
\begin{center}
\includegraphics[scale=0.30]{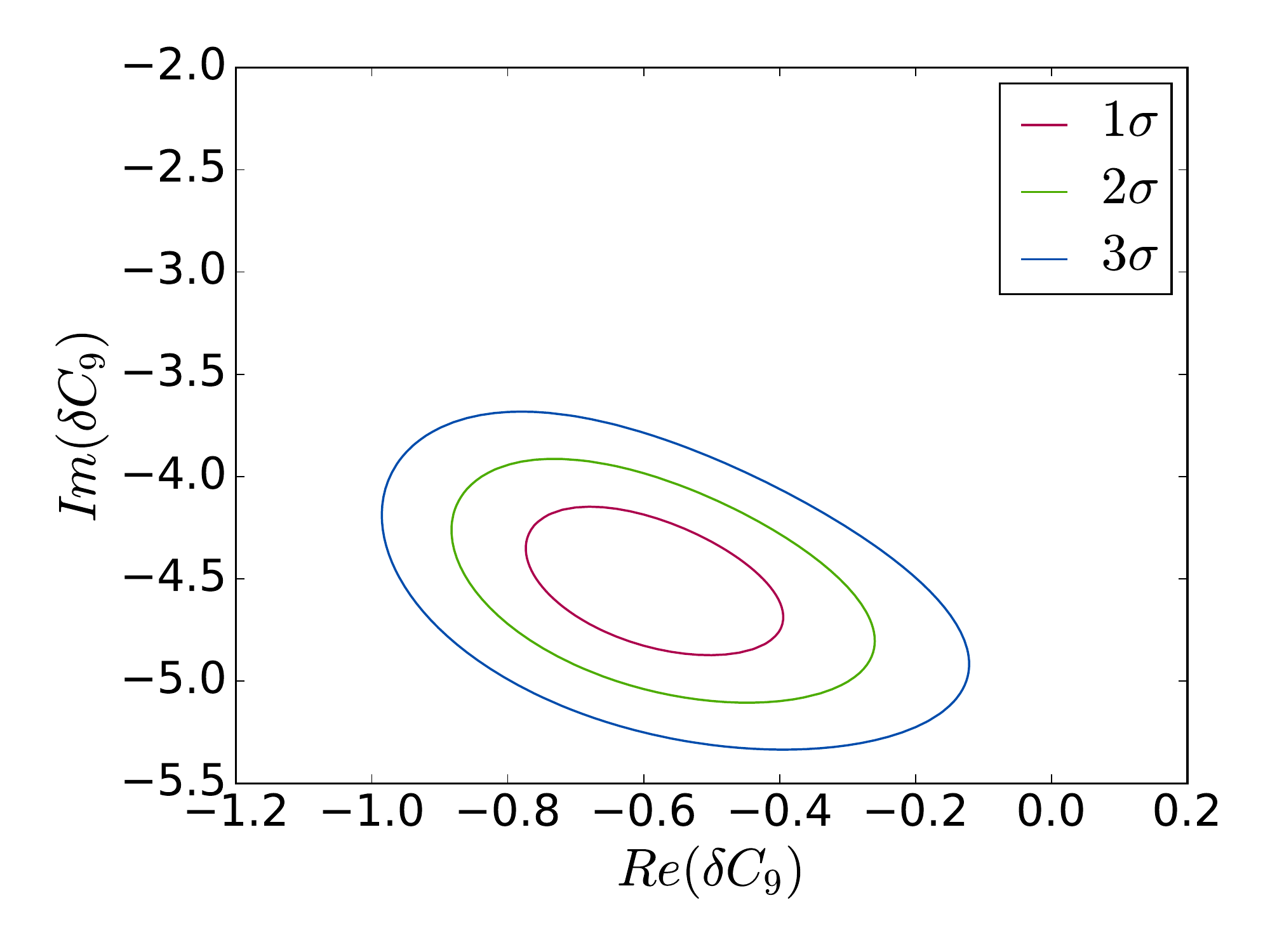} 
\includegraphics[scale=0.30]{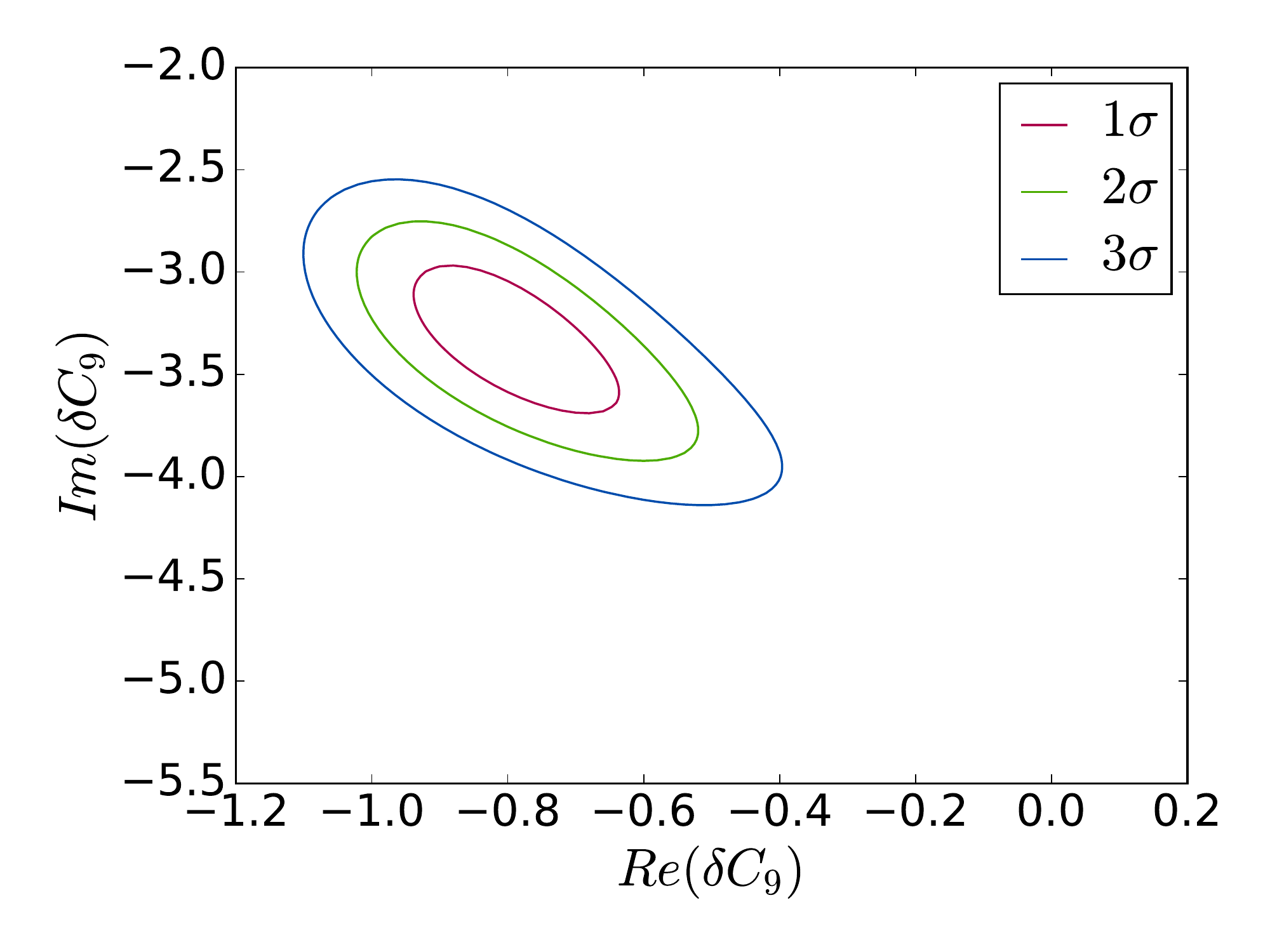}

\caption{Projections of the real and imaginary part of $\delta C_9$ assuming integrated luminosity of 300 fb$^{-1}$. Both $\delta C_9$ and $\delta C_{10}$ are allowed to vary. The contours on the left correspond to fits up to $q^2 = 6$ GeV$^2$, and the contours on the right correspond to fits up to 
$q^2=8$ GeV$^2$. 
 \label{fig:upgrade4}}
\end{center}
\end{figure}

The LHCb detector will be upgraded~\cite{Bediaga:1443882} and is expected to collect a total integrated luminosity of 50 fb$^{-1}$. 
A second upgrade at a high-luminosity LHC will allow for a full dataset of up to 300 fb$^{-1}$. Projections of $C_{9}$ for 300 fb$^{-1}$
are shown in Figs.~\ref{fig:upgrade2} and \ref{fig:upgrade4}, 
where we assumed the current central values and scaled down the present LHCb uncertainties by a factor 10.

Will it be possible to clear the situation with higher statistics and distinguish between NP and hadronic parameter hypotheses?  As we discussed in section 2, within the comparison of the two fits it might be that a larger $q^2$ dependence induced by new data is found which disproves the NP option.  More quantitatively:
Looking at Table~\ref{tab:hParamsFree}, we notice that almost all hadronic parameters are compatible with zero. 
In the hypothetical case that the central values of these parameters stay, while the uncertainties drastically decrease, the LHCb upgrade would strongly favour the hadronic fit. 

On the other hand, it will not be possible to disprove the hadronic hypothesis  in favour of the NP one  with  the present observables as long as the  NP fit is embedded in the more general hadronic fit. But of course  new observables confirming the NP option  can rule out the hadronic hypothesis as discussed before.

%% file: conclusions.tex
\section{Conclusions}\label{sec:conclusions}

In view of the persisting deviations with the SM predictions in the rare $B^0 \to K^{*0} \ell^+\ell^-$ data accumulated by the LHCb experiment during the first run, we address the question of whether these deviations originate from new physics or from unknown large hadronic power corrections by performing global fits to NP in the Wilson coefficients and to unknown power corrections, and doing a statistical comparison.

We showed that the NP fit can be embedded into the hadronic fit what allows for a direct comparison of the two options.

Our analysis shows that -- with the present data -- adding the hadronic parameters does not improve the fit compared to the NP  fit. Hence, our result is a strong indication that the NP interpretation is still a valid option, even if the situation remains inconclusive. 

We discussed the prospects of the statistical comparison of the two hypotheses when the LHCb (upgrade) will offer more statistics in the future.

We reviewed possible options for establishing NP before Belle-II which will be able to resolve the puzzle by measuring the inclusive modes. A confirmation of the deviation in the theoretically clean ratio $R_K$ would indirectly support the NP option in the angular observables.   
And a future estimation of the non-factorisable power corrections using the LCSR approach may 
allow to distinguish the two hypotheses.

\hspace{1cm}

{\bf Note added:}  While finalising the write-up of the results~\cite{Nazilatalk:2016} -- which were presented at the workshop ``Implications of LHCb measurements and future prospects'', CERN, Geneva, 12-14 October 2016 -- another preprint on hadronic uncertainties appeared on the archive~\cite{Capdevila:2017ert}. We add some comments on it: The authors of  Ref.~\cite{Capdevila:2017ert} claim that  factorisable power corrections cannot account for the anomalies. 
In our global analysis of all the  present $b \to s$ data in Ref.~\cite{Hurth:2016fbr}, we show that
the deviation can be reduced by doubling the error in the form factor calculation of Ref.\cite{Straub:2015ica}. 
So one cannot rule out the option that the present anomalies are partially a result of 
underestimated uncertainties in the form factor determination. This finding calls for an independent calculation of the form factors and for further consistency checks with the lattice results.

Moreover, the authors of Ref.~\cite{Capdevila:2017ert} present fits of the non-factorisable power corrections to the data using polynomials with increasing degree  in the three independent helicity amplitudes.
 They  show that there is no statistical significance for a non-trivial $q^2$ dependence beyond the linear order terms  and then conclude that these findings disfavour the option of the non-factorisable power corrections being the explanation for the LHCb anomalies. As we argued (see section 2), the modest $q^2$ dependence of the corrections found in the fits to the present data does not rule out their interpretation as hadronic corrections; for example possible resonances might be smeared out via the large binning (2 GeV$^2$) of the experimental data. 

One also should keep in mind that the NP fit is governed by one Wilson coefficients $C_9$ (or two when including $C_7$ in the NP fit), while their fit with linear polynomials includes six independent coefficients, $h^{(0)}_\lambda$ and $h^{(1)}_\lambda$ ($\lambda=+,-,0$), corresponding to the three helicity amplitudes. The direct comparison of the NP fit with the hadronic fit, as done in the present paper, seems more reasonable in order to get hints for distinguishing the two hypotheses.

%% file: constrained.tex
\section{Fit results assuming $h_+^{(0)}$ to be constrained}\label{app:constrained}
\begin{figure}[h!]
\centering
\vspace*{0.5cm}
\includegraphics[width=0.7\textwidth]{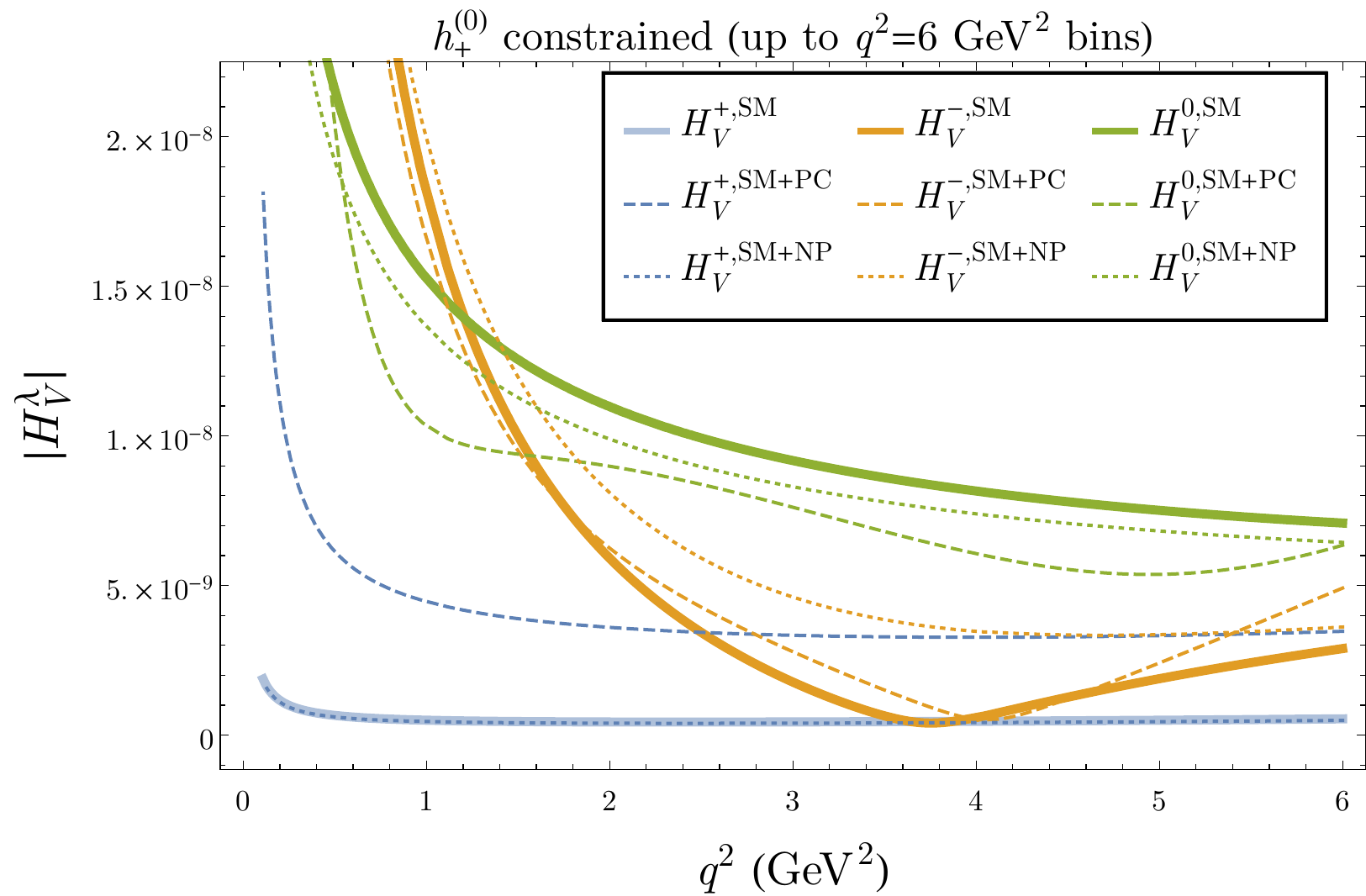}
\includegraphics[width=0.7\textwidth]{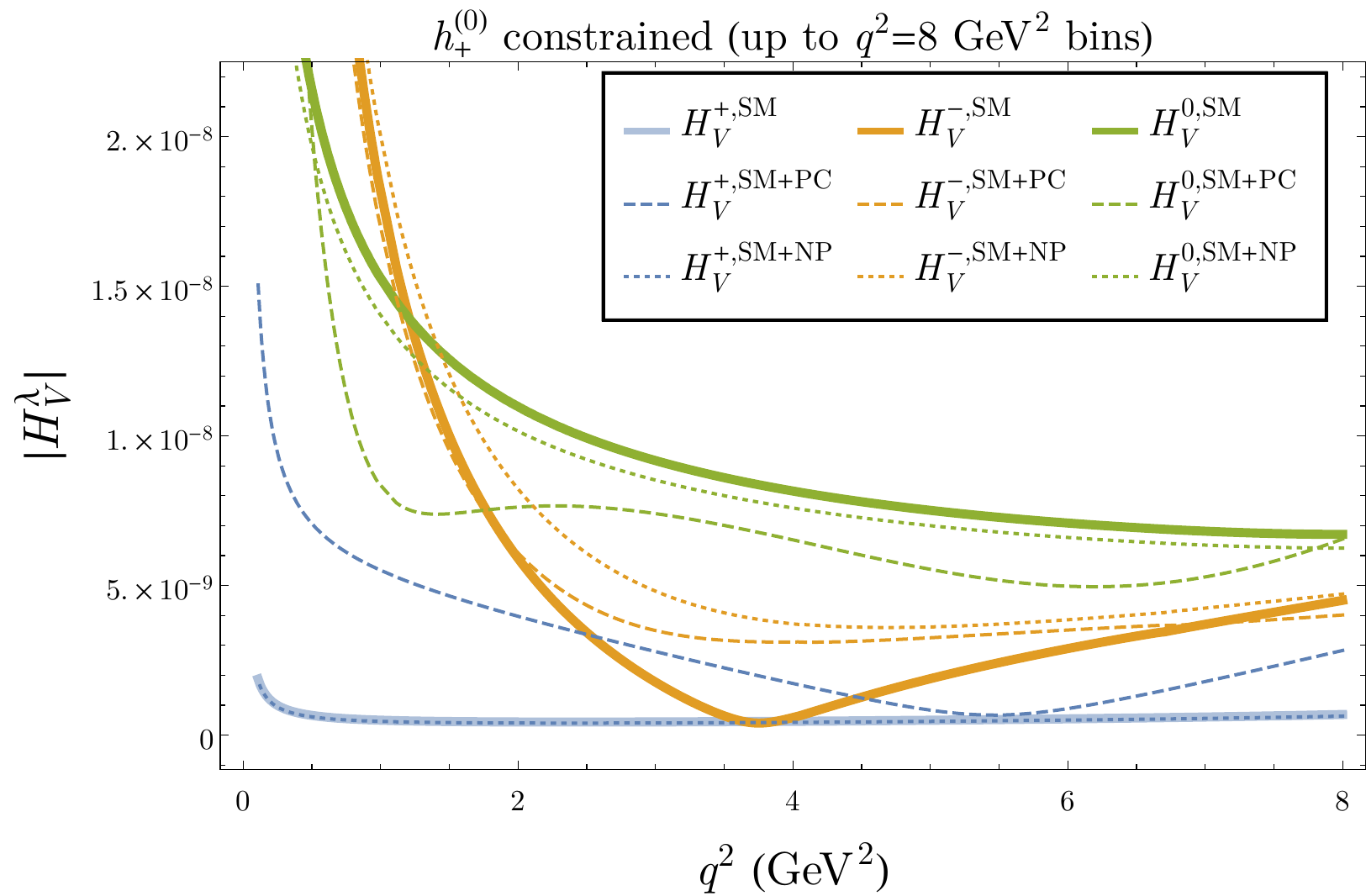}
\caption{Behaviour of absolute value of \textcolor{myblue}{$H_V^+$}, \textcolor{myorange}{$H_V^-$} and \textcolor{mygreen}{$H_V^0$}.
The ``SM'',  ``SM + power correction'' and ``SM + NP'' are shown with solid, dashed and dotted lines, respectively.
Assuming $h_{+}^{(0)}$ to be constrained in the fit for power corrections.
\label{fig:HVallAbsConstrained}}
\end{figure}

The authors of Ref.~\cite{Jager:2012uw} give arguments that the suppression of the helicity amplitude $H_+$ 
with respect to $H_-$ holds also for beyond leading order. In Table~\ref{tab:hParamsConstrained} we have given the power correction fit assuming 
$h_{+}^{(0)}$ to be constrained by $|h_{+}^{(0)}/h_{-}^{(0)}|< 0.2$.
Considering the constraint on $h_{+}^{(0)}$, the effect of the new physics 
and the power corrections fits are compared with SM for $|H_V^\lambda(q^2)|$ in Fig.~\ref{fig:HVallAbsConstrained}.
Here, again the $q^2$ shapes of the power corrections effects are similar 
to the NP effect. 

\begin{table}[h!]
\ra{0.90}
\rb{1.3mm}
\begin{center}
\setlength\extrarowheight{2pt}
\footnotesize{
\begin{tabular}{|l||c|c|}
\hline
 \multicolumn{3}{|c|}{up to $q^2=6$ GeV$^2$ obs.}           \\  
\hline
& Real                          & Imaginary    \\ 
 \hline
$h_+^{(0)}$     & $ ( 7.6 \pm  11.1 )  \times 10^{-5} $ & $ (-6.1 \pm  11.2 ) \times 10^{-5} $ \\
$h_+^{(1)}$     & $ ( 6.8 \pm  12.1 )  \times 10^{-5} $ & $ (-1.6 \pm  2.2  ) \times 10^{-4} $ \\
$h_+^{(2)}$     & $ (-2.2 \pm  3.0  )  \times 10^{-5} $ & $ ( 6.3 \pm  478.5 ) \times 10^{-7}$ \\
\hline 
$h_-^{(0)}$     & $ (-7.6 \pm  18.4 )  \times 10^{-5} $ & $ ( 4.8 \pm  3.2  ) \times 10^{-4} $ \\
$h_-^{(1)}$     & $ (-3.1 \pm  18.5 )  \times 10^{-5} $ & $ (-7.4 \pm  3.3  ) \times 10^{-4} $ \\
$h_-^{(2)}$     & $ ( 2.6 \pm  3.5 )  \times 10^{-5}  $ & $ ( 1.5 \pm  0.6 ) \times 10^{-4} $ \\
\hline 
$h_0^{(0)}$     & $ (-1.6 \pm  3.2 )  \times 10^{-4}  $ & $ ( 7.9 \pm  3.3 ) \times 10^{-4}  $ \\
$h_0^{(1)}$     & $ ( 5.3 \pm  4.0 )  \times 10^{-4}  $ & $ (-1.1 \pm  0.4 ) \times 10^{-3}  $ \\
$h_0^{(2)}$     & $ (-7.6 \pm  6.7 )  \times 10^{-5} $ & $ ( 1.8 \pm  0.9 ) \times 10^{-4}  $ \\
\hline
\end{tabular} 
\begin{tabular}{|l||c|c|}
\hline
 \multicolumn{3}{|c|}{up to $q^2=8$ GeV$^2$ obs.}           \\  
\hline
\rowcolor{white}& Real                          & Imaginary    \\ 
 \hline
$h_+^{(0)}$     & $ ( 4.1 \pm  10.1 ) \times  10^{-5} $ & $ (-4.8 \pm  10.2 ) \times  10^{-5} $ \\
$h_+^{(1)}$     & $ ( 1.9 \pm  1.7 ) \times  10^{-4} $ & $ (-2.4 \pm  2.1 ) \times  10^{-4} $ \\
$h_+^{(2)}$     & $ (-3.6 \pm  2.6 ) \times  10^{-5} $ & $ ( 4.1 \pm  3.3 ) \times  10^{-5} $ \\
\hline 
$h_-^{(0)}$     & $ (-1.0 \pm  1.8 ) \times  10^{-4} $ & $ ( 3.0 \pm  3.5 ) \times  10^{-4} $ \\
$h_-^{(1)}$     & $ ( 3.1 \pm  13.3 ) \times  10^{-5} $ & $ (-3.2 \pm  3.6 ) \times  10^{-4} $ \\
$h_-^{(2)}$     & $ ( 8.5 \pm  18.7 ) \times  10^{-6} $ & $ ( 1.6 \pm  5.6 ) \times  10^{-5} $ \\
\hline 
$h_0^{(0)}$     & $ (-2.9 \pm  3.0 ) \times  10^{-4} $ & $ ( 6.8 \pm  3.5 ) \times  10^{-4} $ \\
$h_0^{(1)}$     & $ ( 7.8 \pm  4.1 ) \times  10^{-4} $ & $ (-9.3 \pm  3.4 ) \times  10^{-4} $ \\
$h_0^{(2)}$     & $ (-9.2 \pm  5.3 ) \times  10^{-5} $ & $ ( 1.1 \pm  0.6 ) \times  10^{-4} $ \\ 
\hline 
\end{tabular} 
}
\caption{Fit parameters  for the power corrections assuming $h_{+}^{(0)}$ to be constrained.
\label{tab:hParamsConstrained}}
\end{center} 
\end{table}

%% file: paperdraft.bbl
\begin{thebibliography}{10}

\bibitem{Aaij:2013qta}
{\scshape LHC}b collaboration, R.~Aaij et~al., \emph{{Measurement of
  Form-Factor-Independent Observables in the Decay $B^{0} \to K^{*0} \mu^+
  \mu^-$}}, \href{http://dx.doi.org/10.1103/PhysRevLett.111.191801}{\emph{Phys.
  Rev. Lett.} {\bf 111} (2013) 191801},
  [\href{http://arxiv.org/abs/1308.1707}{{\tt 1308.1707}}].

\bibitem{Aaij:2015oid}
{\scshape LHC}b collaboration, R.~Aaij et~al., \emph{{Angular analysis of the
  $B^{0}\rightarrow K^{*0}\mu^{+}\mu^{-}$ decay}},
  \href{http://dx.doi.org/10.1007/JHEP02(2016)104}{\emph{JHEP} {\bf 02} (2016)
  104}, [\href{http://arxiv.org/abs/1512.04442}{{\tt 1512.04442}}].

\bibitem{Aaij:2015esa}
{\scshape LHC}b collaboration, R.~Aaij et~al., \emph{{Angular analysis and
  differential branching fraction of the decay $B^0_s\to\phi\mu^+\mu^-$}},
  \href{http://dx.doi.org/10.1007/JHEP09(2015)179}{\emph{JHEP} {\bf 09} (2015)
  179}, [\href{http://arxiv.org/abs/1506.08777}{{\tt 1506.08777}}].

\bibitem{Abdesselam:2016llu}
{\scshape Belle} collaboration, A.~Abdesselam et~al., \emph{{Angular analysis
  of $B^0 \to K^\ast(892)^0 \ell^+ \ell^-$}},  in \emph{{Proceedings, LHCSki
  2016 - A First Discussion of 13 TeV Results: Obergurgl, Austria, April 10-15,
  2016}}, 2016.
\newblock \href{http://arxiv.org/abs/1604.04042}{{\tt 1604.04042}}.

\bibitem{Descotes-Genon:2013wba}
S.~Descotes-Genon, J.~Matias and J.~Virto, \emph{{Understanding the $B\to
  K^*\mu^+\mu^-$ Anomaly}},
  \href{http://dx.doi.org/10.1103/PhysRevD.88.074002}{\emph{Phys. Rev.} {\bf
  D88} (2013) 074002}, [\href{http://arxiv.org/abs/1307.5683}{{\tt
  1307.5683}}].

\bibitem{Altmannshofer:2013foa}
W.~Altmannshofer and D.~M. Straub, \emph{{New physics in $B \to K^*\mu\mu$?}},
  \href{http://dx.doi.org/10.1140/epjc/s10052-013-2646-9}{\emph{Eur. Phys. J.}
  {\bf C73} (2013) 2646}, [\href{http://arxiv.org/abs/1308.1501}{{\tt
  1308.1501}}].

\bibitem{Beaujean:2013soa}
F.~Beaujean, C.~Bobeth and D.~van Dyk, \emph{{Comprehensive Bayesian analysis
  of rare (semi)leptonic and radiative $B$ decays}},
  \href{http://dx.doi.org/10.1140/epjc/s10052-014-2897-0,
  10.1140/epjc/s10052-014-3179-6}{\emph{Eur. Phys. J.} {\bf C74} (2014) 2897},
  [\href{http://arxiv.org/abs/1310.2478}{{\tt 1310.2478}}].

\bibitem{Horgan:2013pva}
R.~R. Horgan, Z.~Liu, S.~Meinel and M.~Wingate, \emph{{Calculation of $B^0 \to
  K^{*0} \mu^+ \mu^-$ and $B_s^0 \to \phi \mu^+ \mu^-$ observables using form
  factors from lattice QCD}},
  \href{http://dx.doi.org/10.1103/PhysRevLett.112.212003}{\emph{Phys. Rev.
  Lett.} {\bf 112} (2014) 212003}, [\href{http://arxiv.org/abs/1310.3887}{{\tt
  1310.3887}}].

\bibitem{Hurth:2013ssa}
T.~Hurth and F.~Mahmoudi, \emph{{On the LHCb anomaly in $B \to
  K^*\ell^+\ell^-$}},
  \href{http://dx.doi.org/10.1007/JHEP04(2014)097}{\emph{JHEP} {\bf 04} (2014)
  097}, [\href{http://arxiv.org/abs/1312.5267}{{\tt 1312.5267}}].

\bibitem{Mahmoudi:2014mja}
F.~Mahmoudi, S.~Neshatpour and J.~Virto, \emph{{$B \to K^{*} \mu^{+} \mu^{-}$
  optimised observables in the MSSM}},
  \href{http://dx.doi.org/10.1140/epjc/s10052-014-2927-y}{\emph{Eur. Phys. J.}
  {\bf C74} (2014) 2927}, [\href{http://arxiv.org/abs/1401.2145}{{\tt
  1401.2145}}].

\bibitem{Jager:2012uw}
S.~J{\"a}ger and J.~Martin~Camalich, \emph{{On $B \to V \ell \ell$ at small
  dilepton invariant mass, power corrections, and new physics}},
  \href{http://dx.doi.org/10.1007/JHEP05(2013)043}{\emph{JHEP} {\bf 05} (2013)
  043}, [\href{http://arxiv.org/abs/1212.2263}{{\tt 1212.2263}}].

\bibitem{Jager:2014rwa}
S.~J{\"a}ger and J.~Martin~Camalich, \emph{{Reassessing the discovery potential
  of the $B \to K^{*} \ell^+\ell^-$ decays in the large-recoil region: SM
  challenges and BSM opportunities}},
  \href{http://dx.doi.org/10.1103/PhysRevD.93.014028}{\emph{Phys. Rev.} {\bf
  D93} (2016) 014028}, [\href{http://arxiv.org/abs/1412.3183}{{\tt
  1412.3183}}].

\bibitem{Ciuchini:2015qxb}
M.~Ciuchini, M.~Fedele, E.~Franco, S.~Mishima, A.~Paul, L.~Silvestrini et~al.,
  \emph{{$B\to K^* \ell^+ \ell^-$ decays at large recoil in the Standard Model:
  a theoretical reappraisal}},
  \href{http://dx.doi.org/10.1007/JHEP06(2016)116}{\emph{JHEP} {\bf 06} (2016)
  116}, [\href{http://arxiv.org/abs/1512.07157}{{\tt 1512.07157}}].

\bibitem{Descotes-Genon:2014uoa}
S.~Descotes-Genon, L.~Hofer, J.~Matias and J.~Virto, \emph{{On the impact of
  power corrections in the prediction of $B \to K^*\mu^+\mu^-$ observables}},
  \href{http://dx.doi.org/10.1007/JHEP12(2014)125}{\emph{JHEP} {\bf 12} (2014)
  125}, [\href{http://arxiv.org/abs/1407.8526}{{\tt 1407.8526}}].

\bibitem{Hurth:2016fbr}
T.~Hurth, F.~Mahmoudi and S.~Neshatpour, \emph{{On the anomalies in the latest
  LHCb data}},
  \href{http://dx.doi.org/10.1016/j.nuclphysb.2016.05.022}{\emph{Nucl. Phys.}
  {\bf B909} (2016) 737--777}, [\href{http://arxiv.org/abs/1603.00865}{{\tt
  1603.00865}}].

\bibitem{Mahmoudi:2016mgr}
F.~Mahmoudi, T.~Hurth and S.~Neshatpour, \emph{{Present Status of $b \to s
  \ell^+ \ell^-$ Anomalies}},  in \emph{{6th Workshop on Theory, Phenomenology
  and Experiments in Flavour Physics: Interplay of Flavour Physics with
  electroweak symmetry breaking (Capri 2016) Anacapri, Capri, Italy, June 11,
  2016}}, 2016.
\newblock \href{http://arxiv.org/abs/1611.05060}{{\tt 1611.05060}}.

\bibitem{Hurth:2014zja}
T.~Hurth and F.~Mahmoudi, \emph{{Signs for new physics in the recent LHCb
  data?}},
  \href{http://dx.doi.org/10.1016/j.nuclphysbps.2015.04.007}{\emph{Nucl. Part.
  Phys. Proc.} {\bf 263-264} (2015) 38--43},
  [\href{http://arxiv.org/abs/1411.2786}{{\tt 1411.2786}}].

\bibitem{Hurth:2014vma}
T.~Hurth, F.~Mahmoudi and S.~Neshatpour, \emph{{Global fits to $b \to
  s\ell\ell$ data and signs for lepton non-universality}},
  \href{http://dx.doi.org/10.1007/JHEP12(2014)053}{\emph{JHEP} {\bf 12} (2014)
  053}, [\href{http://arxiv.org/abs/1410.4545}{{\tt 1410.4545}}].

\bibitem{Hurth:2010tk}
T.~Hurth and M.~Nakao, \emph{{Radiative and Electroweak Penguin Decays of $B$
  Mesons}},
  \href{http://dx.doi.org/10.1146/annurev.nucl.012809.104424}{\emph{Ann. Rev.
  Nucl. Part. Sci.} {\bf 60} (2010) 645--677},
  [\href{http://arxiv.org/abs/1005.1224}{{\tt 1005.1224}}].

\bibitem{Hurth:2007xa}
T.~Hurth, \emph{{Status of SM calculations of $b \to s$ transitions}},
  \href{http://dx.doi.org/10.1142/S0217751X07036476}{\emph{Int. J. Mod. Phys.}
  {\bf A22} (2007) 1781--1795},
  [\href{http://arxiv.org/abs/hep-ph/0703226}{{\tt hep-ph/0703226}}].

\bibitem{Hurth:2003vb}
T.~Hurth, \emph{{Present status of inclusive rare $B$ decays}},
  \href{http://dx.doi.org/10.1103/RevModPhys.75.1159}{\emph{Rev. Mod. Phys.}
  {\bf 75} (2003) 1159--1199}, [\href{http://arxiv.org/abs/hep-ph/0212304}{{\tt
  hep-ph/0212304}}].

\bibitem{Aaij:2014ora}
{\scshape LHC}b collaboration, R.~Aaij et~al., \emph{{Test of lepton
  universality using $B^{+}\rightarrow K^{+}\ell^{+}\ell^{-}$ decays}},
  \href{http://dx.doi.org/10.1103/PhysRevLett.113.151601}{\emph{Phys. Rev.
  Lett.} {\bf 113} (2014) 151601}, [\href{http://arxiv.org/abs/1406.6482}{{\tt
  1406.6482}}].

\bibitem{Alonso:2014csa}
R.~Alonso, B.~Grinstein and J.~Martin~Camalich, \emph{{$SU(2)\times U(1)$ gauge
  invariance and the shape of new physics in rare $B$ decays}},
  \href{http://dx.doi.org/10.1103/PhysRevLett.113.241802}{\emph{Phys. Rev.
  Lett.} {\bf 113} (2014) 241802}, [\href{http://arxiv.org/abs/1407.7044}{{\tt
  1407.7044}}].

\bibitem{Hiller:2014yaa}
G.~Hiller and M.~Schmaltz, \emph{{$R_K$ and future $b \to s \ell \ell$ physics
  beyond the standard model opportunities}},
  \href{http://dx.doi.org/10.1103/PhysRevD.90.054014}{\emph{Phys. Rev.} {\bf
  D90} (2014) 054014}, [\href{http://arxiv.org/abs/1408.1627}{{\tt
  1408.1627}}].

\bibitem{Ghosh:2014awa}
D.~Ghosh, M.~Nardecchia and S.~A. Renner, \emph{{Hint of Lepton Flavour
  Non-Universality in $B$ Meson Decays}},
  \href{http://dx.doi.org/10.1007/JHEP12(2014)131}{\emph{JHEP} {\bf 12} (2014)
  131}, [\href{http://arxiv.org/abs/1408.4097}{{\tt 1408.4097}}].

\bibitem{Altmannshofer:2014rta}
W.~Altmannshofer and D.~M. Straub, \emph{{New physics in $b\to s$ transitions
  after LHC run 1}},
  \href{http://dx.doi.org/10.1140/epjc/s10052-015-3602-7}{\emph{Eur. Phys. J.}
  {\bf C75} (2015) 382}, [\href{http://arxiv.org/abs/1411.3161}{{\tt
  1411.3161}}].

\bibitem{Straub:2015ica}
A.~Bharucha, D.~M. Straub and R.~Zwicky, \emph{{$B\to V\ell^+\ell^-$ in the
  Standard Model from light-cone sum rules}},
  \href{http://dx.doi.org/10.1007/JHEP08(2016)098}{\emph{JHEP} {\bf 08} (2016)
  098}, [\href{http://arxiv.org/abs/1503.05534}{{\tt 1503.05534}}].

\bibitem{Altmannshofer:2015sma}
W.~Altmannshofer and D.~M. Straub, \emph{{Implications of $b\to s$
  measurements}},  in \emph{{Proceedings, 50th Rencontres de Moriond
  Electroweak interactions and unified theories}}, pp.~333--338, 2015.
\newblock \href{http://arxiv.org/abs/1503.06199}{{\tt 1503.06199}}.

\bibitem{Capdevila:2016ivx}
B.~Capdevila, S.~Descotes-Genon, J.~Matias and J.~Virto, \emph{{Assessing
  lepton-flavour non-universality from $B\to K^*\ell\ell$ angular analyses}},
  \href{http://dx.doi.org/10.1007/JHEP10(2016)075}{\emph{JHEP} {\bf 10} (2016)
  075}, [\href{http://arxiv.org/abs/1605.03156}{{\tt 1605.03156}}].

\bibitem{Serra:2016ivr}
N.~Serra, R.~Silva~Coutinho and D.~van Dyk, \emph{{Measuring the Breaking of
  Lepton Flavour Universality in $B\to K^*\ell^+\ell^-$}},
  \href{http://arxiv.org/abs/1610.08761}{{\tt 1610.08761}}.

\bibitem{Khodjamirian:2010vf}
A.~Khodjamirian, T.~Mannel, A.~A. Pivovarov and Y.~M. Wang, \emph{{Charm-loop
  effect in $B \to K^{(*)} \ell^{+} \ell^{-}$ and $B\to K^*\gamma$}},
  \href{http://dx.doi.org/10.1007/JHEP09(2010)089}{\emph{JHEP} {\bf 09} (2010)
  089}, [\href{http://arxiv.org/abs/1006.4945}{{\tt 1006.4945}}].

\bibitem{Khodjamirian:2012rm}
A.~Khodjamirian, T.~Mannel and Y.~M. Wang, \emph{{$B \to K \ell^{+}\ell^{-}$
  decay at large hadronic recoil}},
  \href{http://dx.doi.org/10.1007/JHEP02(2013)010}{\emph{JHEP} {\bf 02} (2013)
  010}, [\href{http://arxiv.org/abs/1211.0234}{{\tt 1211.0234}}].

\bibitem{Nazilatalk:2016}
\emph{Next steps and challenges in global fits for \uppercase{r}un 2}, talk given by
  \uppercase{f} .~\uppercase{m}ahmoudi at the workshop {``Implications of LHCb
  measurements and future prospects''}, 12-14.10.2016, CERN, Geneva, Switzerland. 

\bibitem{Chetyrkin:1996vx}
K.~G. Chetyrkin, M.~Misiak and M.~Munz, \emph{{Weak radiative $B$ meson decay
  beyond leading logarithms}},
  \href{http://dx.doi.org/10.1016/S0370-2693(97)00324-9}{\emph{Phys. Lett.}
  {\bf B400} (1997) 206--219}, [\href{http://arxiv.org/abs/hep-ph/9612313}{{\tt
  hep-ph/9612313}}].

\bibitem{Beneke:2000wa}
M.~Beneke and T.~Feldmann, \emph{{Symmetry breaking corrections to heavy to
  light B meson form-factors at large recoil}},
  \href{http://dx.doi.org/10.1016/S0550-3213(00)00585-X}{\emph{Nucl. Phys.}
  {\bf B592} (2001) 3--34}, [\href{http://arxiv.org/abs/hep-ph/0008255}{{\tt
  hep-ph/0008255}}].

\bibitem{Grinstein:1988me}
B.~Grinstein, M.~J. Savage and M.~B. Wise, \emph{{$B \to X_s e^+ e^-$ in the
  Six Quark Model}},
  \href{http://dx.doi.org/10.1016/0550-3213(89)90078-3}{\emph{Nucl. Phys.} {\bf
  B319} (1989) 271--290}.

\bibitem{Misiak:1992bc}
M.~Misiak, \emph{{The $b \to se^+ e^-$ and $b \to s\gamma$ decays with
  next-to-leading logarithmic QCD corrections}},
  \href{http://dx.doi.org/10.1016/0550-3213(95)00029-R,
  10.1016/0550-3213(93)90235-H}{\emph{Nucl. Phys.} {\bf B393} (1993) 23--45}.

\bibitem{Buras:1994dj}
A.~J. Buras and M.~Munz, \emph{{Effective Hamiltonian for $B \to X_s e^+ e^-$
  beyond leading logarithms in the NDR and HV schemes}},
  \href{http://dx.doi.org/10.1103/PhysRevD.52.186}{\emph{Phys. Rev.} {\bf D52}
  (1995) 186--195}, [\href{http://arxiv.org/abs/hep-ph/9501281}{{\tt
  hep-ph/9501281}}].

\bibitem{Beneke:2001at}
M.~Beneke, T.~Feldmann and D.~Seidel, \emph{{Systematic approach to exclusive
  $B \to V \ell^+ \ell^-, V \gamma$ decays}},
  \href{http://dx.doi.org/10.1016/S0550-3213(01)00366-2}{\emph{Nucl. Phys.}
  {\bf B612} (2001) 25--58}, [\href{http://arxiv.org/abs/hep-ph/0106067}{{\tt
  hep-ph/0106067}}].

\bibitem{Beneke:2004dp}
M.~Beneke, T.~Feldmann and D.~Seidel, \emph{{Exclusive radiative and
  electroweak $b \to d$ and $b \to s$ penguin decays at NLO}},
  \href{http://dx.doi.org/10.1140/epjc/s2005-02181-5}{\emph{Eur. Phys. J.} {\bf
  C41} (2005) 173--188}, [\href{http://arxiv.org/abs/hep-ph/0412400}{{\tt
  hep-ph/0412400}}].

\bibitem{Aaij:2016flj}
{\scshape LHC}b collaboration, R.~Aaij et~al., \emph{{Measurements of the
  S-wave fraction in $B^{0}\rightarrow K^{+}\pi^{-}\mu^{+}\mu^{-}$ decays and
  the $B^{0}\rightarrow K^{\ast}(892)^{0}\mu^{+}\mu^{-}$ differential branching
  fraction}}, \href{http://dx.doi.org/10.1007/JHEP11(2016)047}{\emph{JHEP} {\bf
  11} (2016) 047}, [\href{http://arxiv.org/abs/1606.04731}{{\tt 1606.04731}}].

\bibitem{Mahmoudi:2007vz}
F.~Mahmoudi, \emph{{SuperIso: A Program for calculating the isospin asymmetry
  of $B \to K^* \gamma$ in the MSSM}},
  \href{http://dx.doi.org/10.1016/j.cpc.2007.12.006}{\emph{Comput. Phys.
  Commun.} {\bf 178} (2008) 745--754},
  [\href{http://arxiv.org/abs/0710.2067}{{\tt 0710.2067}}].

\bibitem{Mahmoudi:2008tp}
F.~Mahmoudi, \emph{{SuperIso v2.3: A Program for calculating flavor physics
  observables in Supersymmetry}},
  \href{http://dx.doi.org/10.1016/j.cpc.2009.02.017}{\emph{Comput. Phys.
  Commun.} {\bf 180} (2009) 1579--1613},
  [\href{http://arxiv.org/abs/0808.3144}{{\tt 0808.3144}}].

\bibitem{Lazzaro:2010zza}
A.~Lazzaro and L.~Moneta, \emph{{MINUIT package parallelization and
  applications using the RooFit package}},
  \href{http://dx.doi.org/10.1088/1742-6596/219/4/042044}{\emph{J. Phys. Conf.
  Ser.} {\bf 219} (2010) 042044}.

\bibitem{Wilks:1938dza}
S.~S. Wilks, \emph{{The Large-Sample Distribution of the Likelihood Ratio for
  Testing Composite Hypotheses}},
  \href{http://dx.doi.org/10.1214/aoms/1177732360}{\emph{Annals Math. Statist.}
  {\bf 9} (1938) 60--62}.


\bibitem{CMS:2014xfa}
{LHCb, CMS} collaboration, V.~Khachatryan et~al., \emph{{Observation
  of the rare $B^0_s\to\mu^+\mu^-$ decay from the combined analysis of CMS and
  LHCb data}}, \href{http://dx.doi.org/10.1038/nature14474}{\emph{Nature} {\bf
  522} (2015) 68--72}, [\href{http://arxiv.org/abs/1411.4413}{{\tt
  1411.4413}}].
  
  
\bibitem{Bediaga:1443882}
{\scshape LHC}b collaboration, I.~Bediaga, J.~M. De~Miranda,
  F.~Ferreira~Rodrigues, J.~Magnin, A.~Massafferri, I.~Nasteva et~al.,
  \emph{{Framework TDR for the LHCb Upgrade: Technical Design Report}},  Tech.
  Rep. CERN-LHCC-2012-007. LHCb-TDR-12, Apr, 2012.

\bibitem{Capdevila:2017ert}
B.~Capdevila, S.~Descotes-Genon, L.~Hofer and J.~Matias, \emph{{Hadronic
  uncertainties in $B\to K^*\mu^+\mu^-$: a state-of-the-art analysis}},
  \href{http://arxiv.org/abs/1701.08672}{{\tt 1701.08672}}.

\end{thebibliography}
